\documentclass[twocolumn,tighten]{aastex7}
\usepackage{hyperref}
\usepackage{apjfonts}
\usepackage{apjfonts} 
\usepackage{graphicx} 
\usepackage{amssymb,amsmath} 
\usepackage{color} 

\renewcommand{\dbltextfloatsep}{4mm}
\renewcommand{\textfloatsep}{4mm}

\newcommand{\xrism}{\textit{XRISM}}

\begin{document}

\vspace*{-7.5mm}\hfill{\footnotesize ApJ Letters accepted}
\vspace*{7.5mm}

\righthead{\xrism\ vs simulations comparison}
\lefthead{\xrism\ Collaboration}

\title{\Large Comparing XRISM cluster velocity dispersions with predictions from cosmological simulations: are feedback models too ejective?}

\suppressAffiliations
\correspondingauthor{Nhut Truong (ntruong@umbc.edu) and Skylar Grayson (sigrayso@asu.edu) \\
\hspace*{-4mm}Authors' affilliations are given at the end of the preprint.}

\collaboration{0}{XRISM Collaboration:%
}

\author[0000-0003-4721-034X]{Marc Audard}
\affiliation{Department of Astronomy, University of Geneva, Versoix CH-1290, Switzerland} 
\email{Marc.Audard@unige.ch}

\author{Hisamitsu Awaki}
\affiliation{Department of Physics, Ehime University, Ehime 790-8577, Japan}
\email{awaki@astro.phys.sci.ehime-u.ac.jp}

\author[0000-0002-1118-8470]{Ralf Ballhausen}
\affiliation{Department of Astronomy, University of Maryland, College Park, MD 20742, USA}
\affiliation{NASA / Goddard Space Flight Center, Greenbelt, MD 20771, USA}
\affiliation{Center for Research and Exploration in Space Science and Technology, NASA / GSFC (CRESST II), Greenbelt, MD 20771, USA}
\email{ballhaus@umd.edu}

\author[0000-0003-0890-4920]{Aya Bamba}
\affiliation{Department of Physics, University of Tokyo, Tokyo 113-0033, Japan}
\email{bamba@phys.s.u-tokyo.ac.jp}

\author[0000-0001-9735-4873]{Ehud Behar}
\affiliation{Department of Physics, Technion, Technion City, Haifa 3200003, Israel}
\email{behar@physics.technion.ac.il}

\author[0000-0003-2704-599X]{Rozenn Boissay-Malaquin}
\affiliation{Center for Space Sciences and Technology, University of Maryland, Baltimore County (UMBC), Baltimore, MD, 21250 USA}
\affiliation{NASA / Goddard Space Flight Center, Greenbelt, MD 20771, USA}
\affiliation{Center for Research and Exploration in Space Science and Technology, NASA / GSFC (CRESST II), Greenbelt, MD 20771, USA}
\email{rozennbm@umbc.edu}

\author[0000-0003-2663-1954]{Laura Brenneman}
\affiliation{Center for Astrophysics | Harvard-Smithsonian, Cambridge, MA 02138, USA}
\email{lbrenneman@cfa.harvard.edu}

\author[0000-0001-6338-9445]{Gregory V.\ Brown}
\affiliation{Lawrence Livermore National Laboratory, Livermore, CA 94550, USA}
\email{brown86@llnl.gov}

\author[0000-0002-5466-3817]{Lia Corrales}
\affiliation{Department of Astronomy, University of Michigan, Ann Arbor, MI 48109, USA}
\email{liac@umich.edu}

\author[0000-0001-8470-749X]{Elisa Costantini}
\affiliation{SRON Netherlands Institute for Space Research, Leiden, The Netherlands}
\email{e.costantini@sron.nl}

% ...

\begin{abstract}
The dynamics of the intra-cluster medium (ICM), the hot plasma that fills galaxy clusters, are shaped by gravity-driven cluster mergers and feedback from supermassive black holes (SMBH) in the cluster cores. XRISM measurements of ICM velocities in several clusters offer insights into these processes. We compare XRISM measurements for nine galaxy clusters (Virgo, Perseus, Centaurus, Hydra A, PKS\,0745--19, A2029, Coma, A2319, Ophiuchus)  with predictions from three state-of-the-art cosmological simulation suites, TNG-Cluster, The Three Hundred Project GADGET-X, and GIZMO-SIMBA, that employ different models of feedback. In cool cores, XRISM reveals systematically lower velocity dispersions than the simulations predict, with all ten measurements below the median simulated values by a factor $1.5-1.7$ on average and all falling within the bottom $10\%$ of the predicted distributions. The observed kinetic-to-total pressure ratio is also lower, with a median value of $2.2$\%, compared to the predicted $5.0-6.5$\% for the three simulations. Outside the cool cores and in non-cool-core clusters, simulations show better agreement with XRISM measurements, except for the outskirts of the relaxed, cool-core cluster A2029, which exhibits an exceptionally low kinetic pressure support ($<1\%$), with none of the simulated systems in either of the three suites reaching such low levels. The non-cool-core Coma and A2319 exhibit dispersions at the lower end but within the simulated spread. Our comparison suggests that the three numerical models may overestimate the kinetic effects of SMBH feedback in cluster cores. Additional XRISM observations of non-cool-core clusters will clarify if there is a systematic tension in the gravity-dominated regime as well.  
\end{abstract}


\section{Introduction}

As the most massive gravitationally bound structures in the Universe, galaxy clusters form relatively late in the cosmic timeline, through the hierarchical assembly of smaller structures \citep{kravtsov.borgani.2012}. The space between the member galaxies of a cluster is permeated by the intra-cluster medium (ICM), a tenuous, hot, magnetized X-ray emitting plasma with density $n_H\sim 10^{-3}-10^{-2}$ cm$^{-3}$, temperature $T\sim10^{7-8}$ K, and magnetic field magnitude $B\sim 1\mu$G. This medium is constantly disturbed by large-scale structure formation, driven by accretion and mergers of smaller clusters and groups, which deposit significant amounts of kinetic energy in the form of gas motions. On smaller scales, supermassive black holes (SMBH) located at the centers of brightest cluster galaxies (BGCs) further stir the ICM through feedback activity, introducing additional turbulence and bulk motions \citep[e.g.,][]{fabian2012}. 

Probing these motions has long been challenging for CCD-based X-ray instruments such as Chandra or XMM-Newton, due to their limited spectral resolution (120--150 eV). As a result, the level of ICM motions could only be inferred with very large uncertainties \citep[e.g.,][]{sanders.etal.2020} or indirectly, such as through analyses of fluctuations in X-ray surface brightness maps \citep[e.g.,][]{zhuravleva.etal.2014,heinrich.etal.2024}. The advent of microcalorimeter-based detectors marked a significant breakthrough by improving the resolution by a factor of 20--30. In 2016, the Hitomi mission achieved unprecedented spectral resolution of 5 eV, enabling the first direct measurements of ICM motions in the Perseus cluster \citep{hitomi.collaboration.2016, hitomi.2018}. In 2023, XRISM (X-ray Imaging and Spectroscopy Mission)\footnote{https://xrism.isas.jaxa.jp/en/}, the Hitomi successor, was launched. The mission carries two instruments: Xtend, an X-ray CCD imager, and Resolve, an X-ray microcalorimeter array with a 4.5 eV resolution and a $3'\times3'$ field of view (FoV). Since its launch, XRISM Resolve has delivered high-resolution spectra for a diverse sample of galaxy clusters, including Virgo (XRISM Collaboration et al., submitted), Centaurus \citep{xrism.centaurus.2025}, PKS\,0745--19 (XRISM Collaboration et al., to be submitted), Perseus \citep{xrism.perseus.2025}, A2029 \citep{xrism.a2029.2025a, xrism.a2029.2025b,xrism.a2029.2025c}, A2319 \citep{xrism.a2319.2025}, Coma \citep{xrism.coma.2025}, Hydra-A \citep{rose.etal.2025}, and Ophiuchus \citep{fujita.etal.2025}. These high-resolution spectroscopic observations provide direct measurements of gas motions through line shifts and broadening, which respectively give the ICM average line-of-sight bulk velocity and velocity dispersion (e.g., caused by turbulence). 

From the theoretical side, several numerical studies have been dedicated to predicting and interpreting kinematic measurements from Hitomi and XRISM \citep[e.g.,][]{truong.etal.2024, vazza.brunetti.2025, groth.etal.2025a, groth.etal.2025b}. \cite{truong.etal.2024} performed XRISM Resolve mock observations of a sample of Perseus-like systems drawn from the TNG-Cluster cosmological simulation. Their results showed that the simulated systems exhibit kinetic pressure support levels broadly consistent with those measured by Hitomi, although the observed value lies near the lower end of the simulation predictions. More recently, \cite{groth.etal.2025b} conducted a similar analysis using the SLOW non-radiative simulation, focusing on comparisons with Hitomi/XRISM observations of Perseus and Coma. They also report overall agreement between the simulation predictions and the observed level of kinetic pressure support. In another recent study, \cite{vazza.brunetti.2025} employed an ideal MHD simulation to investigate velocity structures in a Coma-like system and find agreement with the XRISM measurements from \cite{xrism.coma.2025}. 

In this paper, we directly compare XRISM Resolve observations of nine galaxy clusters, seven from the Performance Verification (PV) phase (Virgo, Centaurus, PKS\,0745--19, Perseus, A2029, A2319, and Coma) and two General Observers (GO) targets (Hydra A and Ophiuchus). We contrast these observations with predictions from three state-of-the-art cosmological simulation suites: TNG-Cluster \citep{nelson.etal.2024}, and the Three Hundred Project simulations \citep{cui.etal.2018} using two different physical models, GADGET-X \citep{rasia.etal.2015} and GIZMO-SIMBA \citep{cui.etal.2022}, making every effort to reproduce the XRISM observation setup to ensure direct comparison. 

Our analysis focuses on XRISM's kinematic measurements, specifically velocity dispersions and the inferred kinetic pressure ratio, to validate the physical prescriptions used in the current simulation models. The simulations differ in their treatment of hydrodynamics and baryonic physics--particularly SMBH feedback--and are therefore expected to predict different gas motions in cluster cores. While gravity-driven large-scale structure formation should be implemented very similarly, the resolution and plasma physics implementation differ between the codes and may conceivably result in different predicted ICM kinematics in the gravity-dominated regions as well. 

The paper is organized as follows. Section \ref{sec:data} describes the observational and simulated data, along with key details of the analysis, including the selection of simulated clusters analogous to the observed ones. Section \ref{sec:result} presents the main results, focusing on key comparisons between XRISM observations and simulation predictions. In Section \ref{sec:discussion}, we discuss the implications of these findings for current simulation models. Finally, Section \ref{sec:summary} summarizes the key results.  

\section{Data}
\label{sec:data}
\subsection{Observations} 

All XRISM observations of the cluster sample were conducted either during the PV phase or shortly thereafter, within GO Cycle 1. Table \ref{tab:sample} summarizes the full list of targets, individual pointings, and corresponding references. For each source, spectra were extracted from the full field of view of the Resolve instrument (except a partial field was used for Centaurus, PKS\,0745--19, and Ophiuchus) and plasma spectral lines were modeled using the atomic transition database AtomDB version 3.0.9 or 3.1.3 \citep{smith.etal.2001}. The spectral modeling employed absorbed single-temperature models (except for Centaurus's central pointing) of optically thin plasma in collisional ionization equilibrium (tbabs*bvapec), accounting for non-X-ray background, AGN emission in cool core clusters, resonant scattering when necessary (Perseus, A2029, PKS0745-19, which slightly reduces the inferred velocity dispersions for Perseus and PKS0745-19), and thermal broadening of the spectral lines. Most of the velocity constraints come from the dominant Fe He$\alpha$ line complex at $E\simeq6.7$ keV (rest frame). For each spectrum, the line-of-sight (LOS) velocity dispersion is derived with respect to the mean LOS velocity given by the line centroid. No attempts were made (nor is it possible in observations) to separate the contribution of any large-scale LOS velocity variations from the velocity dispersion, which thus includes variations on all scales \citep[for discussion see][]{xrism.coma.2025}. The resulting best-fit velocity dispersions and temperatures are provided in Table \ref{tab:sample}. For details on the modeling procedures, we refer the reader to the specific publications cited in the table.

\begin{table*}[tb]
\centering
\label{table:fits}
\renewcommand{\tabcolsep}{3mm}
\small\noindent
\caption{XRISM measurements of PV and GO clusters. The quoted uncertainties in observed velocity dispersion, temperature, and derived kinetic pressure ratio correspond to $1\sigma$. For the halo mass ($M_{\rm 200c}$), the quoted uncertainties represent the mass range used in selecting the observed analogs.} 
\hspace*{-1.5cm}
\begin{tabular}{p{3.7cm}cccccccc}
\hline\hline
 &\multicolumn{1}{c}{$M_{\rm 200c}$} &  \multicolumn{1}{c}{Velocity dispersion}    & \multicolumn{1}{c}{Temperature} &  \multicolumn{1}{c}{$P_{\rm kin}/P_{\rm total}$} & \multicolumn{1}{c}{Reference}\\
& [$M_\odot$]  & [km/s] & [keV] \\
\hline
Virgo Center (CC) \dotfill & $10^{14.4\pm0.15}$ & $153^{+16}_{-16}$ & $1.802^{+0.041}_{-0.040}$ & $0.074\pm0.011$ & XRISM Collaboration et al. (submitted) \\
 & & & & & \cite{simionescu.etal.2017} \\
Virgo NW \dotfill & & $61.8^{+16}_{-18}$ &  $2.240^{+0.069}_{-0.063}$ & $0.010\pm0.004$ & XRISM Collaboration et al. (submitted)\\ 
Centaurus (CC)$^+$ \dotfill & $10^{14.6\pm0.15}$ & $117^{+9}_{-9}$ & $2.11^{+0.28}_{-0.28}$ & $0.038\pm0.007$ & \cite{xrism.centaurus.2025}\\
 & & & & & \cite{reiprich.bohringer.2002} \\ 
Hydra-A (CC) \dotfill & $10^{14.73\pm0.15}$ & $160^{+10}_{-11}$ & $3.34^{+0.21}_{-0.17}$ & $0.045\pm0.005$ & \cite{rose.etal.2025}\\
 & & & & & \cite{girardi.etal.2022} \\ 
PKS0745-19 (CC) \dotfill & $10^{14.81\pm0.15}$ & $121^{+17}_{-17}$ & $5.866^{+0.098}_{-0.097}$ & $0.015\pm0.003$ & XRISM Collaboration et al. (to be submitted)\\
 & & & & & \cite{george.etal.2009} \\ 
Perseus C0 (CC)$^*$ \dotfill & $10^{14.85\pm0.15}$ & $153^{+7}_{-6}$ & $3.82^{+0.07}_{-0.05}$ & $0.036\pm0.002$ & \cite{xrism.perseus.2025}\\ 
 & & & & & \cite{simionescu.etal.2011} \\
Perseus C1$^*$ \dotfill & & $101^{+10}_{-9}$ & $4.79^{+0.08}_{-0.08}$ & $0.013\pm0.002$ & \cite{xrism.perseus.2025}\\ 
Perseus M1$^*$ \dotfill &  & $179^{+20}_{-19}$ & $6.28^{+0.27}_{-0.16}$ & $0.031\pm0.005$ & \cite{xrism.perseus.2025}\\ 
Perseus O1$^*$ \dotfill & & $184^{+32}_{-29}$ & $7.67^{+0.36}_{-0.44}$ & $0.027\pm0.007$ & \cite{xrism.perseus.2025}\\ 
A2029 Center (CC) \dotfill & $10^{14.9\pm0.15}$ & $148^{+13}_{-9}$ & $6.62^{+0.11}_{-0.11}$ & $0.020\pm0.003$ & \cite{xrism.a2029.2025b}\\
 & & & & & \cite{walker.etal.2012} \\ 
A2029 N1 \dotfill & & $58^{+37}_{-48}$ & $7.49^{+0.21}_{-0.21}$ & $0.003\pm0.003$ & \cite{xrism.a2029.2025b}\\ 
A2029 N2 \dotfill & & $94^{+44}_{-50}$ & $8.26^{+0.34}_{-0.31}$ &$0.007\pm0.004$ & \cite{xrism.a2029.2025b}\\ 
Ophiuchus Inner (CC) \dotfill &$10^{15.22\pm0.15}$  & $115^{+7}_{-7}$ & $5.8^{+0.2}_{-0.2}$ &$0.014\pm0.001$ & \cite{fujita.etal.2025}\\
 & & & & & \cite{giacintucci.etal.2020} \\ 
Ophiuchus Outer \dotfill && $186^{+9}_{-9}$ & $8.4^{+0.2}_{-0.2}$ &$0.025\pm0.002$ & \cite{fujita.etal.2025}\\ 
A2319 (NCC) \dotfill & $10^{15.02\pm0.15}$ & $232^{+14}_{-14}$ & $9.48^{+0.20}_{-0.20}$ & $0.034\pm0.003$ & \cite{xrism.a2319.2025}\\
 & & & & & \cite{ghirardini.etal.2018} \\ 
Coma Center (NCC)\dotfill & $10^{15.25\pm0.15}$ & $208^{+12}_{-12}$ & $8.55^{+0.25}_{-0.25}$ & $0.030\pm0.004$ &  \cite{xrism.coma.2025}\\
 & & & & & \cite{ho.etal.2022} \\ 
Coma South \dotfill & & $202^{+24}_{-24}$ & $7.44^{+0.44}_{-0.44}$ &$0.033\pm0.006$ &  \cite{xrism.coma.2025}\\
\hline
\begin{minipage}{20cm}
$^{+}$Central-region measurements; temperature is the mean of the two-component fit.\\
$^*$Full-FoV measurements were provided via private communication.
\end{minipage}
\end{tabular}
 
\label{tab:sample}
\end{table*}

\subsection{Cosmological simulations}
\label{sec:codes}

We compare the XRISM observations with quantitative predictions extracted from three recent cosmological simulation suites of galaxy clusters: TNG-Cluster, Three Hundred Project (the 300 hereafter) GADGET-X and SIMBA-GIZMO. The three suites each produce large samples of a few hundred massive galaxy clusters ($M_{\rm 200c}\gtrsim10^{14}M_\odot$)\footnote{$M_{\rm 200c}$ is defined as the total mass enclosed within $R_{200c}$, the radius inside which the mean density is 200 times the universe's critical density ($\rho_c$). Explicitly, $M_{\rm 200c}=4/3\pi 200\times \rho_c R_{\rm 200c}^3$.} within a $\Lambda$CDM universe with similar cosmological parameters \citep{planck.collaboration.2016}. 

\begin{itemize}
    \item {\bf TNG-Cluster} is a suite of cosmological magneto-hydrodynamical simulations of galaxy clusters \citep{nelson.etal.2024}. It is an extension of the IllustrisTNG series of cosmological simulations of galaxy formation and evolution: TNG50, TNG100, and TNG300 \citep{marinacci.etal.2018, naiman.etal.2018, nelson.etal.2018, nelson.etal.2019a, nelson.etal.2019b,pillepich.etal.2018b, springel.etal.2018}. TNG-Cluster comprises 352 massive clusters with $M_{\rm 200c}\gtrsim10^{14.5}M_\odot$, which are drawn from a 1 Gpc Dark Matter (DM) simulated box. 
    \item {\bf The 300 GADGET-X} suite models the formation and evolution of 324 clusters with the halo mass $M_{\rm 200c}\gtrsim10^{14.8}M_\odot$ \citep{cui.etal.2018} at z=0. These clusters are extracted from the DM-only simulation \citep{klypin.etal.2016}. The hydrodynamical re-simulations are performed with the modern Smooth-Particle-Hydrodynamics (SPH) code GADGET-X \citep{beck.etal.2016} with high-fidelity baryonic physics \citep{rasia.etal.2015}.
    \item {\bf The 300 SIMBA-GIZMO} suite models the same 324 clusters as in the 300 GADGET-X simulations but with a different hydrodynamical scheme and baryonic model \citep{cui.etal.2022}. It employs the GIZMO numerical scheme \citep{hopkins.etal.2015} and the SIMBA model of galaxy formation \citep{dave.etal.2019}.
\end{itemize}

The three sets of simulations incorporate different models of baryonic physics, and all account for key physical processes relevant to the formation and evolution of galaxies: radiative cooling and heating from a UV/X-ray radiation field, star formation, metal enrichment, stellar feedback, black hole seeding, growth, and feedback. Because these processes occur at scales smaller than those resolved by large-scale cosmological simulations, they are modeled in a ``subgrid'' manner and coupled with resolved hydrodynamics and gravity to account for their effects within an evolving cosmological context. However, these three simulations differ in numerical resolution, numerical methods, and the implementation of SMBH feedback.  

In terms of numerical resolution, TNG-Cluster has a baryon mass resolution of $\sim1.7\times10^{7}M_\odot$, which is $\sim 20$x better than that in the Three Hundred runs ($\sim3.5\times10^{8}M_\odot$). In all three simulations,  the spatial resolution in cluster cores ($<100$ kpc) ranges from a few kpc up to $\sim10$ kpc, which is well below the FoV of the core pointings ($\gtrsim60$ kpc). At larger radii ($\sim500$ kpc) relevant for the outermost observed regions in A2029, the effective resolution is $\gtrsim20$ kpc, which is also smaller than the FoV of the A2029 offset pointings ($>200$ kpc).

TNG-Cluster implements SMBH feedback using a two-mode scheme, in which the feedback energy is released into the surrounding medium either in thermal form, when the SMBH accretion rate is high relative to the Eddington limit, or in kinetic form, when the accretion rate is low \citep{weinberger.etal.2017,pillepich.etal.2018b}. In both cases, the energy is distributed in intermittent, randomly oriented wind, that is isotropic when averaged over many discrete outbursts in time. In addition, the radiation emitted by accreting SMBHs impacts the local gas thermodynamics.

The SIMBA-GIZMO run also adopts a multiple-mode feedback depending on the SMBH accretion rate. At high accretion rates, feedback is released as kinetic wind (AGN wind mode), while at low accretion rates it is released in a jet mode that can drive collimated outflows with velocities of $\sim10^4$ km/s. In the jet mode, there is an additional heating from X-ray scattering off the accretion disc. In both modes, the kinetic energy is injected along the polar directions with respect to the inner accretion disc \citep{dave.etal.2019}. 

The GADGET-X run employs a single-mode feedback prescription, with the feedback energy released isotropically and purely in thermal form \citep{rasia.etal.2015}.

\subsection{Selecting simulated analogs of observed clusters}
\label{sec:selection}

\begin{figure}
\centering
\vspace*{2mm}
\includegraphics[width=8.5cm]{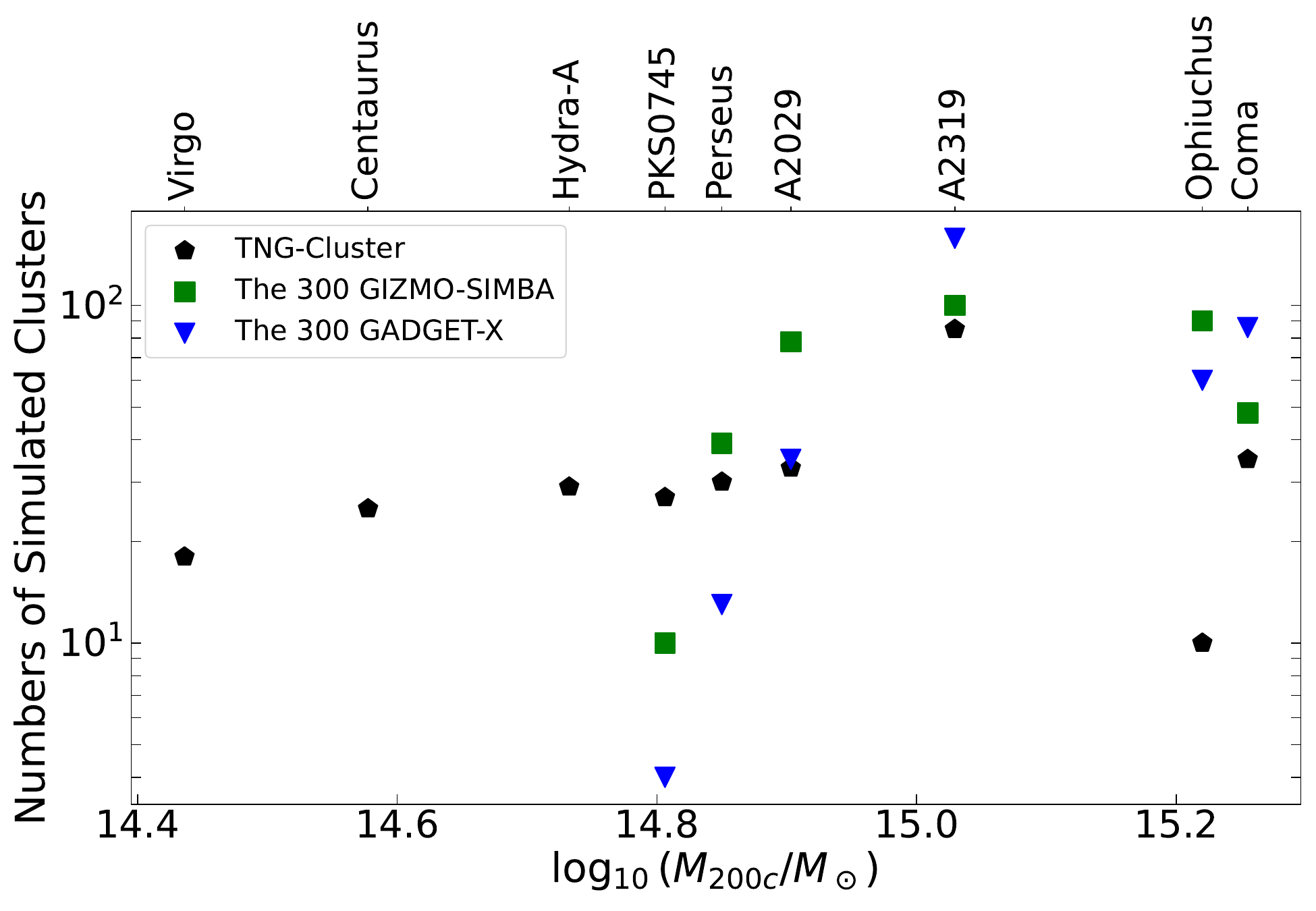}
\caption{The number of simulated analogs of the observed clusters contained in the three simulation suites: TNG-Cluster, The 300 GADGET-X and GIZMO-SIMBA, as a function of the halo mass ($M_{\rm 200c}$). The analogs are selected based on the observed halo mass and core properties (see \S\ref{sec:selection}).}
\label{fig:mass_dis}
\end{figure}

Simulated clusters span a wide range of masses and dynamic properties, from relaxed systems to current mergers, with or without dense cool cores and a central SMBH. It is important to compare XRISM velocity dispersions to simulated clusters with similar global properties. For that, we select the analogs of the observed clusters based on two criteria:

\begin{itemize}
    \item {\bf Halo mass.} For each observed target, we select simulated clusters with halo masses ($M_{\rm 200c}$) within a $\pm0.15$ dex of the observed value, as summarized in Table \ref{tab:sample}. We do not attempt to match the exact mass values reported in the literature, as these estimates are subject to uncertainties, particularly those arising from differences in measurements across various X-ray instruments \citep{schellenberger.etal.2015}. Instead, we adopt a broad mass bin of $\pm0.15$ dex ($\sim40\%$) to define the analog selection.   

    \item {\bf Core properties.} We classify the simulated clusters as cool-core and non cool-core according to their central cooling time defined as
    \begin{equation}
        t_{\rm cool} = \frac{3}{2}\frac{(n_e+n_i)k_BT}{n_en_i\Lambda}, \label{eqn:tcool}
    \end{equation}
    where $n_i$ is the ion number density, $n_e$ the electron number density, T the temperature of each gas element, $\Lambda$ the cooling function, and $k_B$ the Boltzmann constant. The central cooling time is computed using average quantities within the radius of $0.015R_{\rm500c}$ \citep{mcdonald.etal.2013,lehle.etal.2024}. We adopt an observation-motivated threshold of $t_{\rm cool}<1$ Gyr to identify cool-core (CC) systems \citep{mcdonald.etal.2013}. The rest are non-cool-core (NCC). To test the robustness of our results with respect to CC/NCC classification, we also apply an alternative proxy, the central entropy ($K=k_B T/n_e^{2/3}$). The comparisons presented in the followings remain consistent when using this entropy-based criterion.
\end{itemize}

Figure~\ref{fig:mass_dis} presents the number of observed analogs predicted by the three simulations. The 300 project includes clusters with masses from that of our  PKS\,0745--19 ($M_{\rm 200c}\sim10^{14.8}M_\odot$) and above. The TNG-Cluster suite includes masses down to the upper range of the Virgo cluster ($M_{\rm 200c}\sim10^{14.4}M_\odot$).

\begin{figure*}
\centering
\vspace*{2mm}
\includegraphics[width=18.0cm]{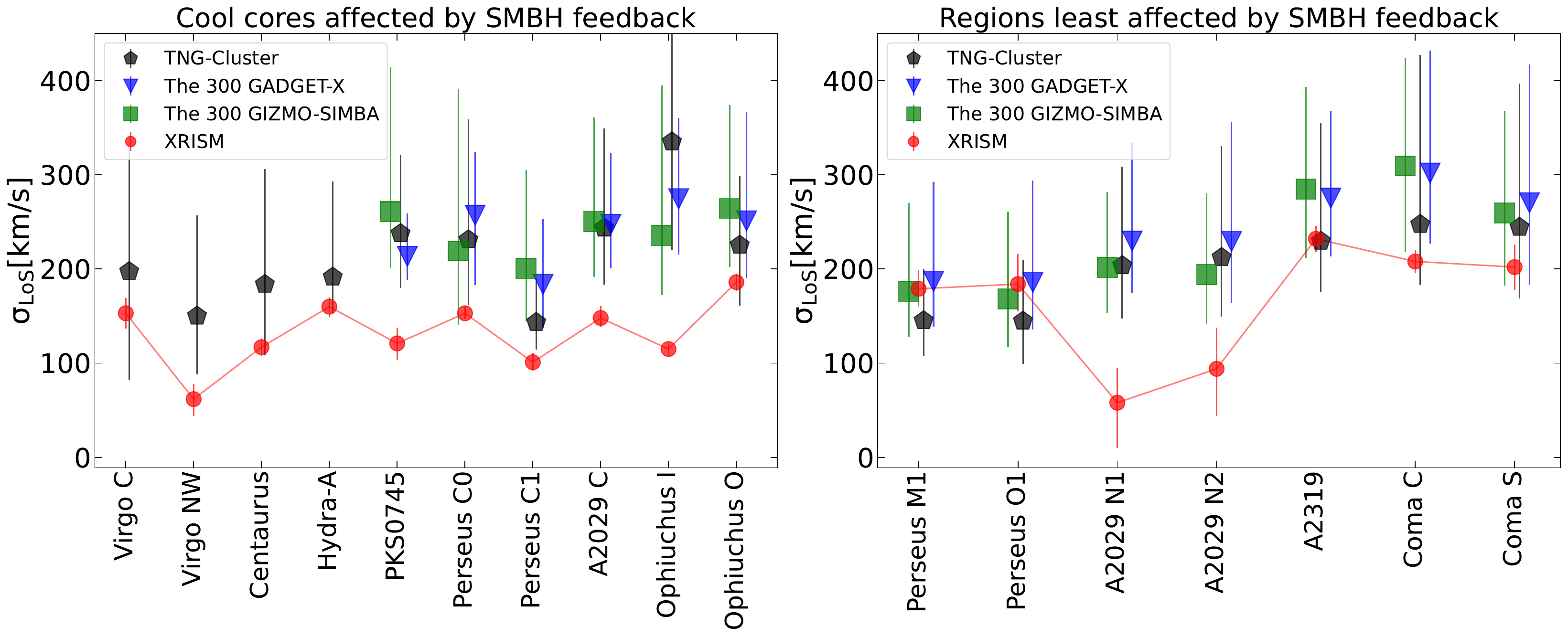}
\caption{Comparison between XRISM observations and simulation predictions of the velocity dispersion. {\it Left panel} displays the result for the regime where SMBH feedback should be significant or possibly dominant. This includes central pointings in CCs, the Perseus C1 and the Virgo NW offset pointings, which are within their cooling radius ($\lesssim100$ kpc). {\it Right panel} shows the result for the regime  least affected by SMBH feedback, including central pointings in NCC clusters and the offset pointings outside the cool cores: Perseus M1 and O1, A2029 N1 and N2. The error bars for the XRISM points represent statistical measurement uncertainties. The simulation points show the median for the corresponding simulated sample, while the error bars show the 68\% range of values in the sample ($16^{\rm th}-84^{\rm th}$ percentile envelope). The observed velocity dispersions are systematically lower than the simulation medians, in particular for the SMBH feedback dominated regime.}
\label{fig:sigma}
\end{figure*}

\subsection{Computation of simulated quantities}

To compare with XRISM observations, we compute emission-weighted quantities: the line-of-sight velocity dispersion ($\sigma_{\rm LoS}$) and the emission-weighted temperature ($T_{\rm ew}$), defined as
\begin{equation}
    \sigma_{\rm LoS} = \sqrt{\frac{\Sigma_i w_i {\rm v}_{\rm LoS,i}^2}{\Sigma_i w_i}-\bigg(\frac{\Sigma_i w_i {\rm v}_{\rm LoS,i}}{\Sigma_i w_i}\bigg)^2},
\end{equation}
\begin{equation}
    T_{\rm ew} = \frac{\Sigma_i w_i T_{\rm i}}{\Sigma_i w_i},
\end{equation}
where ${\rm v}_{\rm LoS,i}$ is the velocity of the i-th gas cell along a random line-of-sight, $T_{i}$ is its temperature, and $w_i$ is its X-ray emission in the [6.0-8.0] keV energy band. The summation is performed over all gas cells within a rectangular volume along the LoS, with a projected area corresponding to the physical XRISM Resolve FoV of the observed target. The X-ray emission is computed using the gas cells' density, temperature, and metallicity, assuming an APEC emission model with atomic data taken from AtomDB, which is implemented in the XSPEC package \citep[version 3.0.9;][]{smith.etal.2001}. We removed cells with $T<10^{5.3}K$ to avoid the effects of dense clumps that may be unreliably modeled in these suites. In appendix A, we examine the consistency between the emission-weighted quantities with those derived from spectral fitting (as done in real observations) for both a CC and a NCC system. The comparison shows agreement within uncertainties, indicating that either approach yields compatible results for comparison with XRISM data. This finding is consistent with the results reported in \cite{truong.etal.2024} for a simulated sample of Perseus-like clusters. 

The observed XRISM velocity dispersions include all motions on the line of sight, without attempting to separate bulk motions from turbulence by filtering variations on large linear scales (which is not possible using the current data). Our dispersion estimates for the simulated clusters do not apply any such filtering either, to be directly comparable with observations. 

To increase the simulated sample, we compute simulated quantities along three random orthogonal lines of sight for each central pointing. While this does not result in a sample of statistically independent dispersion values (because in the case of a well-established isotropic turbulence, there is a correlation between the dispersions along the 3 axes for a given cluster), it does increase our chance of not missing a higher or a lower dispersion along certain directions in dynamically young clusters (e.g., the direction of a recent merger). It is a conservative approach for our qualitative comparison that does not require a statistical sample. 

For each offset pointing, we selected six random locations at the same projected distances to the cluster center as observed, and at each location, we performed computations along two random orthogonal directions. In the following, we report the results for each observed analog by treating all associated random lines of sight and offset locations as a representative population for that analog.   

\section{Results}
\label{sec:result}
In this Section we present the comparisons between XRISM thermal and kinematic measurements of the ICM and predictions from the three sets of simulations. To facilitate this comparison, we divide the analyses into two distinct regimes probed by XRISM observations, dominated by different physical processes: the cool-core and non-cool-core. The cool-core regime includes the central pointings of all seven cool-core clusters, as well as the offset C1 pointing in Perseus, the NW pointing in Virgo, and the outer region of Ophiuchus. Those regions lie within the cooling radius, defined as the radius at which the cooling time equals the Universe's age at the cluster redshift \citep[][]{peres.etal.1998,cavagnolo.etal.2009}. In this regime, the physical state of the ICM is expected to be strongly affected by feedback activity from the central supermassive black hole \citep[as in Perseus,][]{xrism.perseus.2025}. Whereas, the non-cool-core regime reflects ICM conditions primarily driven by cluster growth through subcluster accretion and mergers. This regime is probed by two NCC clusters (Coma and A2319) and offset observations of the cool-core clusters that point beyond their cool-core radii: Perseus M1 and O1, A2029 N1 and N2. By comparing XRISM observations and simulation predictions in these two regimes, we aim to validate the implementation of gravitational and non-gravitational processes in the simulation codes.

\begin{figure*}
\includegraphics[width=18.0cm]{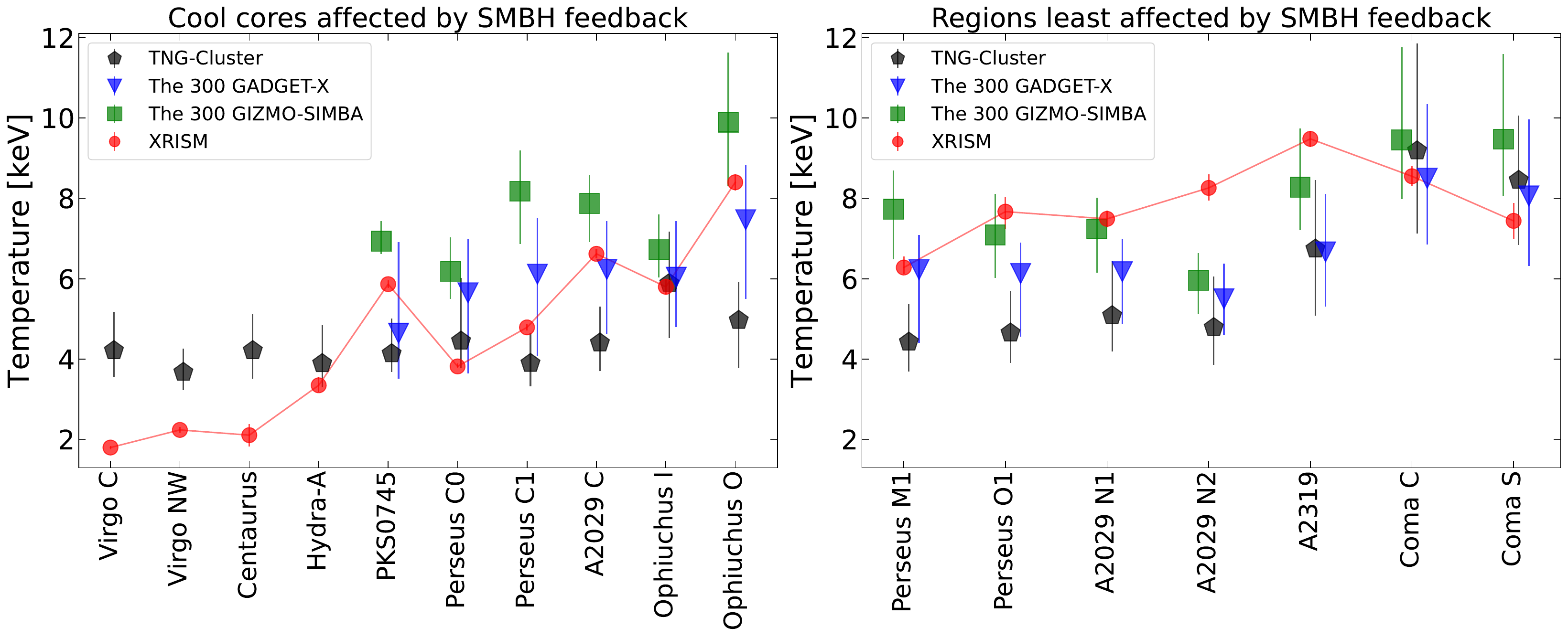}
\caption{Comparison between XRISM observations and simulations predictions of the gas temperature. The descriptions are similar to those in Figure~\ref{fig:sigma}.}
\label{fig:kT}
\end{figure*}

\begin{figure*}
\includegraphics[width=18.0cm]{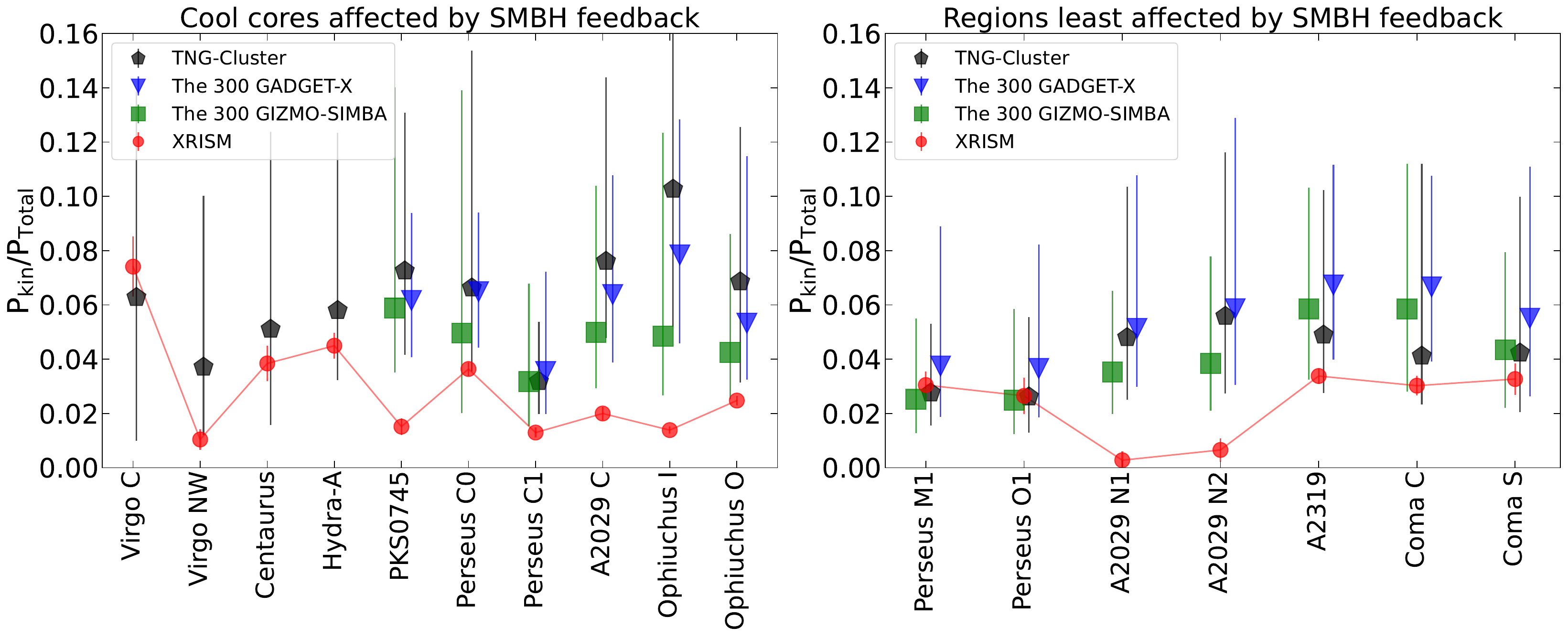}

\caption{Comparison between XRISM observations and simulations predictions of the kinetic pressure ratio. The descriptions are similar to those in Figure~\ref{fig:sigma}.}
\label{fig:nontherm}
\end{figure*}

\subsection{Velocity dispersions}
\label{sec:vels}

We first compare the line-of-sight velocity dispersions. Figure~\ref{fig:sigma} presents the comparison for the two regimes: cool-core ({\it left} panel) and non-cool-core ({\it right} panel). In the cool-core regime, predictions from the three simulations are consistent with one another for systems where all models have observed analogs. However, the measured velocity dispersion values are systematically lower than the simulations predictions. Relative to the TNG-Cluster distributions, the median percentile of the ten observed dispersions is the lower $10\%$. For six of those that also have analogs in the Three Hundred suites, the discrepancy between simulations and observations is similar. While each observation does not represent a particularly unlikely deviation on its own, {\em all ten}\/ observed cool-core pointings fall well below the simulation medians by a factor of 1.5-1.7 on average. 

In the non-cool-core regime, we find a similar level of agreement among the simulation predictions. Overall, the simulations show better agreement with the XRISM data in this regime. Relative to the TNG-Cluster distributions, the median percentile of the observed best-fit dispersions is the lower $\sim30\%$. A significant discrepancy arises for the two offset pointings in A2029 (N1 and N2) --- in none of the three simulation suites we find  any clusters with velocity dispersions as low as those observed in A2029. This will be discussed in more detail below.  

\subsection{Temperature}

Figure~\ref{fig:kT} presents the comparison of gas temperature between XRISM Resolve observations and simulation predictions within the same observed FoV for the two regimes. Unlike for the velocity dispersions where the codes agree, the three simulation suites disagree in the predicted temperatures, most significantly for the feedback-dominated regions.  The Three Hundred GIZMO-SIMBA model consistently predict the highest temperature values, while TNG-Cluster predicts the lowest values. A notable feature for TNG-Cluster for the cool-core regime is a constant median predicted gas temperature ($T\sim4$ keV) across clusters spanning a range of masses. At any fixed mass, however, there is cluster-to-cluster scatter spanning $\sim3-6$ keV. This result suggests that within cool-core regions, the TNG-Cluster predicted gas temperatures are driven more by the interplay between non-gravitational processes such as radiative cooling and SMBH feedback than by gravitational scaling with the cluster mass.

When compared with XRISM temperature measurements, the comparison picture is less straightforward than in the case of velocity dispersion. The level of agreement between simulations and observations varies depending on both the chosen model and the cluster regime. In the cool-core regime, at the high-mass end (comparable to or above the mass of PKS 0745-19), the observed temperatures align more closely with GADGET-X predictions, while TNG-Cluster tends to underpredict and GIZMO-SIMBA systematically overpredicts the measurements. At the low-mass end, where only TNG-Cluster provides predictions, the simulations yield a nearly constant temperature of $\sim4$ keV, as discussed above, which is significantly higher than the observed value of $\sim2$ keV. In the non-cool-core regime, GIZMO-SIMBA shows the best overall consistency with XRISM data, whereas TNG-Cluster and GADGET-X generally underpredict the observed values, with the exception of the two Coma pointings where the agreement is notably improved.

\subsection{Kinetic-to-thermal pressure ratio}

Figure \ref{fig:nontherm} shows the ratio of kinetic to total pressure calculated following 
\begin{equation}
    \frac{P_{\text{kin}}}{P_{\text{total}}} = \frac{1}{1+\left( \frac{k_B T}{\mu m_p \sigma_{\text{LOS}}^2}\right)}
\end{equation}
where $\mu$ is the mean molecular weight and $m_p$ is the proton mass. This expression assumes isotropic gas motions, namely the total velocity dispersion is given by $\sigma_{\rm Total}^2 =3\sigma_{\rm LoS}^2$. Here we find that the relationship is primarily driven by the velocity dispersion, following similar trends to those in Figure \ref{fig:sigma}. In the cool-core regime, the simulations are consistent with each other and they systematically overpredict the XRISM observed level of kinetic pressure support. Except for the Virgo central pointing, all nine other pointings have observed values lying below the simulation averages, with many below the 68\% ranges of the simulated distributions. The median kinetic pressure ratios predicted by TNG-Cluster, GADGET-X, and GIZMO-SIMBA are $6.5_{-2.1}^{+1.0}\%$, $6.3_{-1.3}^{+1.0}\%$, and $5.0_{-1.0}^{+0.3}\%$, respectively, compared to the observed median value of $2.2_{-1.0}^{+2.0}\%$. GIZMO-SIMBA has the lowest values of the pressure ratio, closer to the observations, which is a reflection of the fact that that code produces the highest gas temperatures in the cores, as discussed above. 

In the NCC regime, we find overall a better alignment between simulations and observations; the observed values are below the simulation medians but most are within their 68\% spread. As with the velocity dispersion, the two offset pointings of A2029 (N1 and N2) are clear exceptions: the observed values are significantly below the simulation averages. The observed kinetic pressure ratios for these offset pointings are exceptionally low, $0.3\%$ for N1 and $\sim0.7\%$ for N2, compared to the predicted values of $\gtrsim4\%$. 

\cite{xrism.coma.2025} already noted that both Coma pointings exhibit the kinetic pressure ratios below the simulation predictions. Our GADGET-X value for the Coma central pointing is lower than the \cite{sayers.etal.2021} value from the same code shown in \cite{xrism.coma.2025}, which reduces the discrepancy between GADGET-X and XRISM from factor 3 to factor 2. The change likely arises because in \cite{sayers.etal.2021}, the kinetic pressure ratio was derived from the cluster mass distribution analysis, while here we use the simulation velocity data directly.

\section{Discussion}
\label{sec:discussion}

\subsection{Verifying models of SMBH feedback in cluster cores}

The three simulations analyzed in this study adopt different subgrid models for the effects of SMBH feedback on the surrounding ICM (\S\ref{sec:codes}). The model parameters are calibrated to reproduce the observed galaxy populations, in particular, the galaxy luminosity function \citep{pillepich.etal.2018b, cui.etal.2018, dave.etal.2019}. The large-scale effects of SMBH feedback on the ICM can be broadly classified into two types: thermal and ejective \citep{zinger.etal.2020}. The thermal effect refers to the heating of the gas caused by feedback processes, which has been extensively constrained at the cluster population level using X-ray observations, particularly through mass-temperature scaling relations \citep[e.g.,][]{mahdavi.etal.2013, lieu.etal.2016, truong.etal.2018}. The ejective effect involves the removal and redistribution of gas from the cluster core to larger radii. This process has been more challenging to constrain with previous X-ray instruments, because gas velocities were not accessible.

The kinematic measurements of XRISM in cool-core clusters provide a direct test of ejective effects of SMBH feedback in the simulation models. The comparison of velocity dispersion in Figure~\ref{fig:sigma} indicates that SMBH feedback in all three simulations may drive somewhat more ejective effects than suggested by XRISM observations. This discrepancy appears largely independent of the specific feedback implementation --- whether the energy is deposited in kinetic or thermal form and whether it is injected isotropically or along the SMBH jet axis. This finding suggests a possible tension between the simulated feedback strength and the measured gas motions in cluster cool cores. Future studies (e.g., Truong et al., in prep) comparing radial profiles of velocity dispersions from simulations with spatially resolved XRISM measurements within cool core regions, such as those available for Virgo, Perseus, and Ophiuchus, may provide insight on what pieces of the current feedback models are deficient.

It is possible that the discrepancy in ICM velocities between the simulations and XRISM observations in cool cores reflect incomplete modeling of the ICM physics in the simulations. For instance, none of the three models currently includes non-thermal high-energy particles, or cosmic rays (CRs), known to coexist with the thermal plasma in clusters. CRs can channel some of the SMBH output energy into ICM heating, bypassing the ICM motions, and CRs can also be accelerated by the ICM motions \citep[e.g.,][]{guo.oh.2008,fujita.ohira.2013,brunetti.jones.2014,jacob.pfrommer.2017,yang.etal.2019,kunz.etal.2022}, absorbing some of the ICM kinetic energy.

\begin{figure*}
\includegraphics[width=8.5cm]{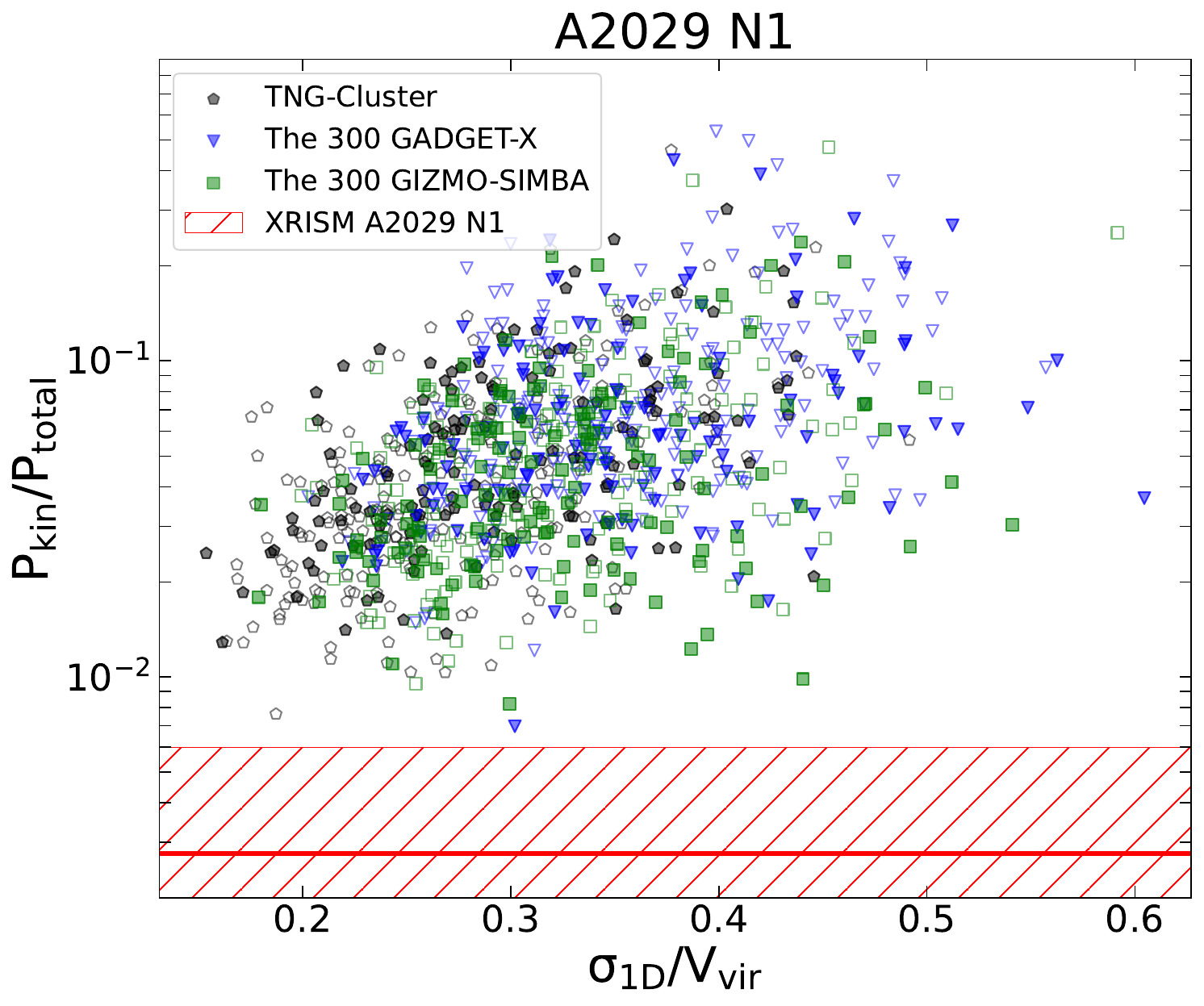}
\includegraphics[width=8.5cm]{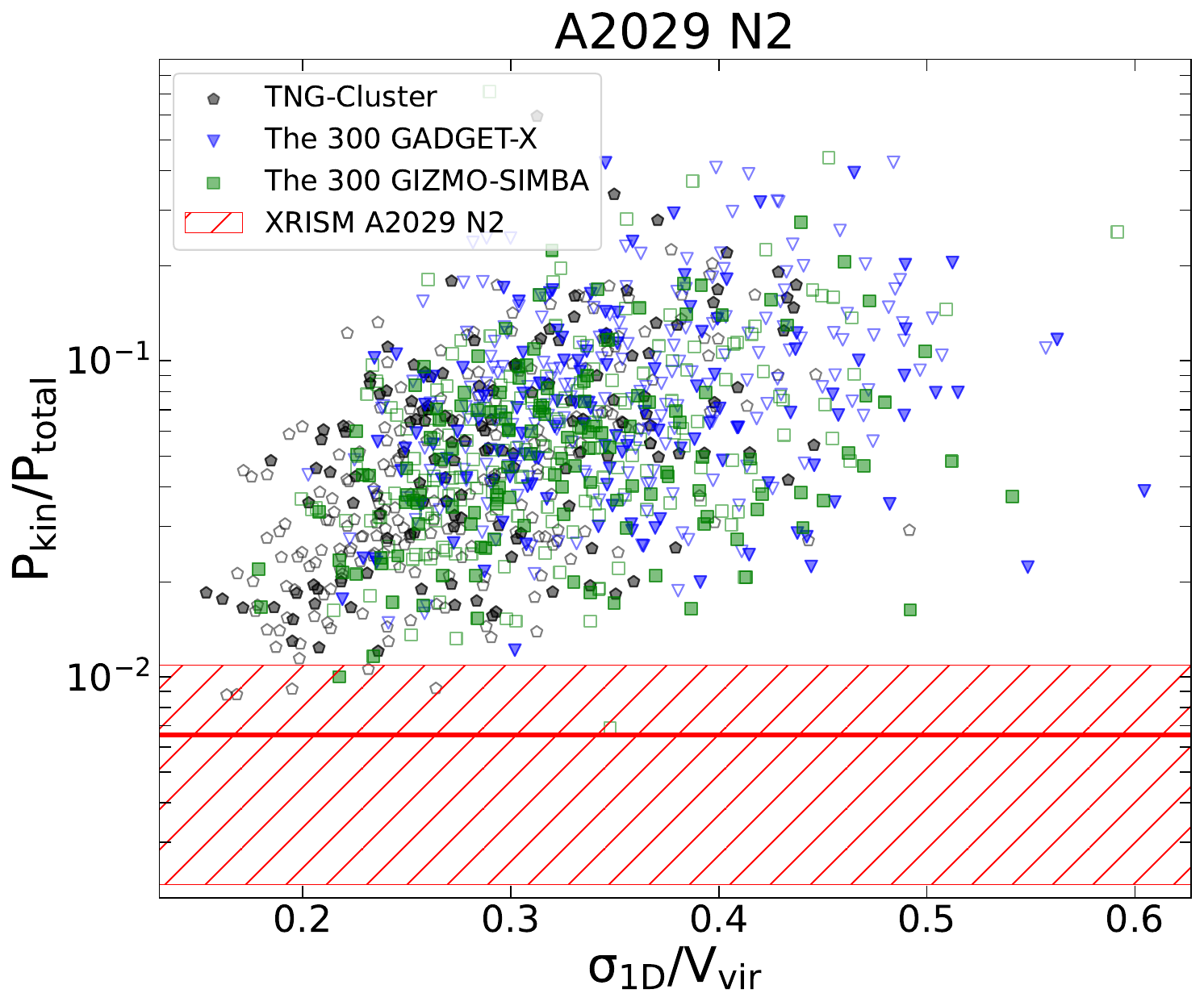}
\caption{The extreme low values of kinetic pressure ratios for A2029 offset pointings N1 ({\it left}) and N2 ({\it right}) in comparison with predictions from the three simulations. The results are shown for 352 simulated clusters at z=0 in TNG-Cluster, 324 clusters in the 300 GADGET-X and GIZMO-SIMBA, where the thermal and kinetic measurements are obtained in a self-similar way to account for the mass effect (see text). The x-axis shows a proxy for the cluster dynamical state, with larger values corresponding to more disturbed clusters (see text for the detailed definition). Filled symbols signify systems with halo mass within the A2029 mass range ($M_{\rm 200c}=10^{14.9\pm0.15}M_\odot$). The solid lines show the medians of the XRISM measurements, and the shaded regions indicate the corresponding $1\sigma$ uncertainties.}
\label{fig:A2029}
\end{figure*}

\subsection{The gravity-dominated regime and the special case of the relaxed cluster A2029}

In regions where SMBH feedback is not expected to strongly influence the ICM's physical state, such as at large distances from cool cores (e.g., Perseus M1, O1) or in non cool-core clusters (A2319 and Coma), the simulations generally show better agreement with XRISM observations of the level of random gas motions. For the two non cool-core clusters, A2319 and Coma, the observed velocity dispersions and the inferred kinetic pressure support are lower than median values for most of the simulations, but typically within the 68\% spread of values in the simulated samples. XRISM observations of a larger sample of non-cool-core systems are needed to see whether there is a systematic discrepancy with simulations for this regime. 

In contrast, the two offset pointings in A2029 (N1 and N2) show strikingly low velocity dispersion and kinetic pressure support compared to simulation predictions. The observed kinetic pressure support is at the sub-percent level. While this discrepancy may reflect shortcomings in modeling of the ICM physics within the simulations, an issue that will require additional XRISM observations of offset pointings to confirm, it could also be due to the limited numbers of A2029-like systems in the simulations, which may not sample rare, extremely relaxed clusters like A2029. The cluster was selected for an early XRISM observation specifically as a cluster with the most relaxed X-ray morphology among the nearby systems, in order to probe the floor of the kinetic pressure support. To further investigate the latter possibility, we expand the A2029-like samples to include all simulated systems from the three simulations and examine how the kinetic pressure support level varies with the cluster's dynamical state. 

Figure \ref{fig:A2029} compares the kinetic pressure ratio in A2029's offset pointings with a total of 1000 simulated clusters from the three simulation suites. The plots display $P_{\rm kin}/P_{\rm total}$ as a function of a proxy for the cluster's dynamical state, defined as the ratio of the averaged isotropic velocity dispersion to the virial velocity:  $\sigma_{\rm 1D}/V_{\rm vir}$, in which $\sigma_{\rm 1D}$ is given by
\begin{equation}
    \sigma_{\rm 1D} = \sqrt{\frac{1}{3}(\bar{\sigma}_{x}^2+\bar{\sigma}_{y}^2+\bar{\sigma}_{z}^2)}, \label{eqn:sigma1D}
\end{equation}
where $\bar{\sigma}_{x,y,z}$ is the mass-weighted velocity dispersion along each direction, computed for gas cells within the core-excise region $[0.1-1]R_{\rm 200c}$. The virial velocity defined as
\begin{equation}
V_{\rm vir}=\sqrt{\frac{G M_{\rm 200c}}{R_{\rm 200c}}}. \label{eqn:vir}
\end{equation}
This proxy characterizes the cluster's dynamical state with lower values indicating more relaxed systems. To account for mass dependence, the kinetic pressure ratio is computed over a FoV of $0.2R_{\rm 500c}\times 0.2R_{\rm 500c}$ and at galactocentric distances of $0.2R_{\rm 500c}$ for N1, and $0.4R_{\rm 500c}$ for N2. These self-similar numbers match the XRISM Resolve FoV and physical distances of the two offset pointings in A2029.

For the N1 pointing, nearly all simulated clusters exhibit kinetic pressure support above $1\%$ level, and even the most relaxed clusters do not reach the observed level of $\sim0.3\%$. In fact, none of the 1000 systems across the three simulation suites reaches the observed N1 value of $P_{\rm kin}/P_{\rm total}+1\sigma$, indicating that the probability of finding such a cluster in the simulations is below $0.1\%$. For N2, a few simulated systems do lie near the observed level of $\sim0.7\%$, though the majority still show significantly higher levels of kinetic pressure support at that radius, with the probability of finding a match in simulations only $0.7\%$ (i.e., $<1\%$). These extremely low probabilities highlight both the rarity of A2029 and the challenge for these models to reproduce the exceptionally low velocity dispersions observed outside its cool core. 

\section{Summary}
\label{sec:summary}

In this paper, we compare XRISM observations of nine galaxy clusters: Virgo, Perseus, Centaurus, Hydra A, PKS\,0745--19, A2029, Coma, A2319, and Ophiuchus, with predictions from three sets of state-of-the-art cosmological simulations: TNG-Cluster, the 300 GADGET-X and GIZMO-SIMBA. For this comparison, we selected clusters from the simulations matching the real clusters based on halo mass and core properties, and computed the simulated velocity dispersions, ICM temperatures, and kinetic pressure ratios within regions matched to the XRISM pointings. The observed pointings are classified into two regimes: i) cool-core regions, which are expected to be strongly influenced by SMBH feedback, and ii) offset pointings outside of the cool core, as well central and offset pointings in non-cool-core clusters, which should be least affected by SMBH feedback and dominated by the effects of cluster mergers. Our main findings are summarized as follows: 

\begin{itemize}
    \item All three simulation suites systematically overpredict XRISM measurements of line-of-sight velocity dispersions in cool-core regions. For all ten cool-core pointings, the observed velocity dispersions fall below the simulation medians for all three suites --- by a factor of 1.5-1.7 on average and all falling within the bottom $\sim10\%$ of the simulated distributions. This result suggests that the studied SMBH feedback models, despite having very different implementations, are too ejective and rely too much on stirring of the ICM to balance the radiative cooling.
  
     \item Simulations also overpredict velocity dispersions in some of the non-cool-core pointings, while giving consistent predictions for some others. Our observation sample of these regions is too small to draw conclusions on the presence of a systematic discrepancy.

    \item While the three suites agree among themselves in their prediction of velocity dispersions, there is systematic disagreement in predicted temperatures between simulations, with the GIZMO-SIMBA runs predicting the highest temperatures and TNG-Cluster runs predicting the lowest. The alignment of XRISM observations with these predictions vary, with low mass clusters generally falling below the TNG predictions, and higher mass clusters falling towards the middle of the three simulations. 

    \item The kinetic to total pressure ratios follow a trend similar to the velocity dispersions, with simulations systematically predicting higher values than observed in cool-core regions. The median kinetic-to-total pressure ratios predicted for cool cores are $6.5_{-2.1}^{+1.0}\%$ (TNG-Cluster), $6.3_{-1.3}^{+1.0}\%$ (GADGET-X), and $5.0_{-1.0}^{+0.3}\%$ (GIZMO-SIMBA), in comparison with the XRISM median value of $2.2_{-1.0}^{+2.0}\%$. Outside of cool-core regions and in non-cool-core clusters, simulations align better with observations (with the notable exception of A2029, see below). We note, however, that this ratio is calculated by combining the velocity dispersion and temperature, and conflates the discrepancies for those two observables. The velocity dispersion and temperature separately should be more informative for testing the simulations.

    \item The simulation suites used in this work fail to capture extreme environments, such as the highly relaxed cluster A2029. The two pointings outside the A2029 cool core exhibit very low levels of kinetic pressure support ($<1\%$). Such low values are not found in any cluster from any the three simulations outputs. Reproducing rare clusters such as A2029 will require either extremely large simulation volumes or targeted zoom simulations, but it can illuminate the ICM dynamics in the absence of recent disturbances. 
\end{itemize}

Our work presents an extensive comparison between recent XRISM cluster observations and cosmological simulations. The clusters observed span a range of masses and dynamic properties, providing a test for simulation models. Our results indicate that the SMBH feedback implemented in current simulations does not fully capture the ICM dynamics in the cluster cool cores. The study also highlights the need for additional XRISM observations of non cool-core clusters, which are essential for better constraining models of the ICM physics in the gravity-dominated regime.

\begin{acknowledgments}
The findings reported here reflect more than thirty years of work by scientists and engineers who developed an X-ray microcalorimeter array and overcame major setbacks. We  thank the entire \xrism\ team for their work in building, launching, calibrating, and operating the observatory. Our thanks also go to the referee for helpful comments. Part of this work was supported by the U.S.\ Department of Energy by Lawrence Livermore National Laboratory under Contract DE-AC52-07NA27344, and by NASA under contracts 80GSFC21M0002 and 80GSFC24M0006 and grants 80NSSC20K0733, 
80NSSC18K0978, 80NSSC20K0883, 80NSSC20K0737, 80NSSC23K0646, 80NSSC24K0678, 80NSSC18K1684, 80NSSC23K0650, and 80NNSC22K1922. Support was provided by JSPS KAKENHI grant numbers JP23H00121, JP22H00158, JP23H04899, JP21K13963, JP24K00638, JP24K17105, JP21K13958, JP21H01095, JP23K20850, JP24H00253, JP21K03615, JP24K00677, JP20K14491, JP23H00151, JP19K21884, JP20H01947, JP20KK0071, JP23K20239, JP24K00672, JP24K17104, JP24K17093, JP20K04009, JP21H04493, JP20H01946, JP23K13154, JP19K14762, JP20H05857, JP23K03459, and JP25H00672, the JSPS Core-to-Core Program, JPJSCCA20220002, and the Strategic Research Center of Saitama University. This work has been made possible by the The Three Hundred collaboration. We acknowledge The Red Española de Supercomputación for granting computing time for running the hydrodynamic simulations of The Three Hundred galaxy cluster project in the Marenostrum supercomputer at the Barcelona Supercomputing Center. We would like to thank Arif Babul, Elena Rasia, Thomas Hough, and Annalisa Pillepich for their helpful discussions during the development of this work. SG acknowledges support from NSF award 2233001.
LC acknowledges support from NSF award 2205918. 
CD acknowledges support from STFC through grant ST/T000244/1. 
LG acknowledges support from Canadian Space Agency grant 18XARMSTMA.
NO acknowledges partial support by the Organization for the Promotion of Gender Equality at Nara Women's University. 
MS acknowledges support by the RIKEN Pioneering Project Evolution of Matter in the Universe (r-EMU) and Rikkyo University Special Fund for Research (Rikkyo SFR). 
AT acknowledges support from the Kagoshima University postdoctoral research program (KU-DREAM). 
SY acknowledges support by the RIKEN SPDR Program. 
IZ acknowledges partial support from the Alfred P.\ Sloan Foundation through the Sloan Research Fellowship.
DN acknowledges funding from the Deutsche Forschungsgemeinschaft (DFG)
through an Emmy Noether Research Group (grant number NE 2441/1-1).
CZ was supported by the GACR grant 21-13491X. S.E. acknowledges the financial contribution from the contracts
Prin-MUR 2022 supported by Next Generation EU (M4.C2.1.1, n.20227RNLY3 {\it The concordance cosmological model: stress-tests with galaxy clusters}), and from the {\it Bando INAF per la Ricerca Fondamentale 2024} with a {\it Theory Grant} on
``Constraining the non-thermal pressure in galaxy clusters with high-resolution X-ray spectroscopy'' (1.05.24.05.10). WC is supported by Atracci\'{o}n de Talento Contract no. 2020-T1/TIC19882 granted by the Comunidad de Madrid and by the Consolidación Investigadora grant no. CNS2024-154838 granted by the Agencia Estatal de Investigación (AEI) in Spain. He also thanks the Ministerio de Ciencia e Innovaci\'{o}n (Spain) for financial support under Project grant PID2021-122603NB-C21, ERC: HORIZON-TMA-MSCA-SE for supporting the LACEGAL-III Latin American Chinese European Galaxy Formation Network) project with grant number 101086388, and the science research grants from the China Manned Space Project, CMS-CSST-2025-A04.

\end{acknowledgments}

\appendix

\section{Comparison between emission-weighted versus spectral-fitting quantities}
\label{app:spec_vs_ew}

To assess how well emission-weighted quantities approximate those obtained from spectral fitting, as done in real X-ray analyses, we perform mock X-ray analysis on two sets of simulated clusters from TNG-Cluster: Perseus-like (CC) and A2319-like (NCC) systems. We follow the end-to-end pipeline for analyzing synthetic X-ray data as presented in \cite{truong.etal.2024}, which consists of two main steps. In the first step, we generate XRISM Resolve mock X-ray observations for central pointings using the PyXSIM package \citep{zuhone.etal.2014} for central pointings. Mock X-ray photons are produced assuming an APEC plasma model with atomic data from AtomDB \citep[version 3.0.9][]{smith.etal.2001}. These photons are subjected to the Doppler effect, due to both thermal and bulk gas motions, as well as galactic absorption modeled using the "wabs" model. The resulting photon flux is then convolved with XRISM Resolve instrumental responses with a Gate Valve-closed ARF and assuming a spectral resolution of 5~eV. This step produces event files that closely resemble real XRISM observations. In the second step, we extract X-ray spectra from the mock event files and perform spectral fitting in the energy band of [6-8] keV to obtain best-fit values of temperature and velocity dispersion. 

\begin{figure*}
\includegraphics[width=17.0cm]{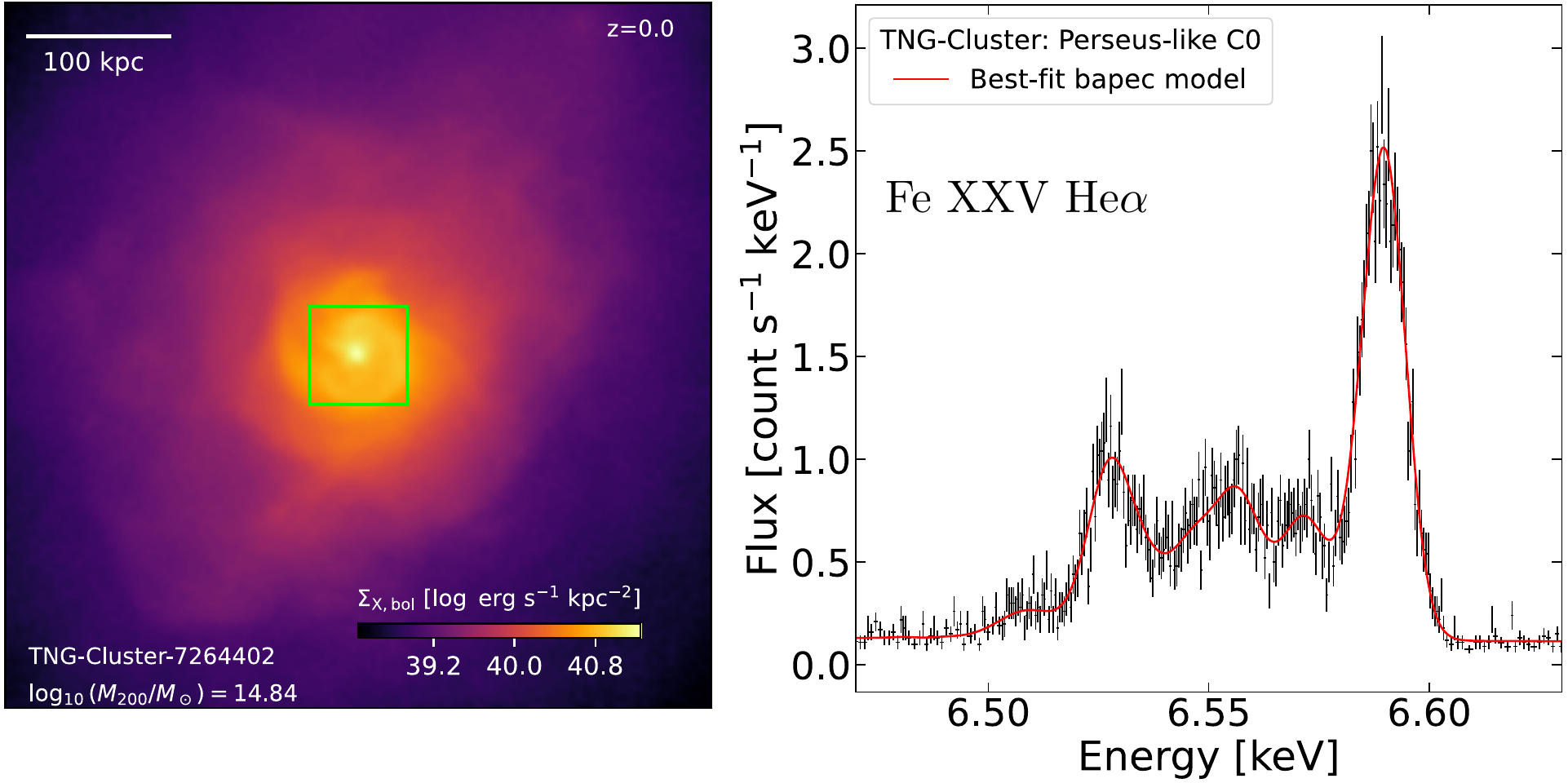}
\includegraphics[width=17.0cm]{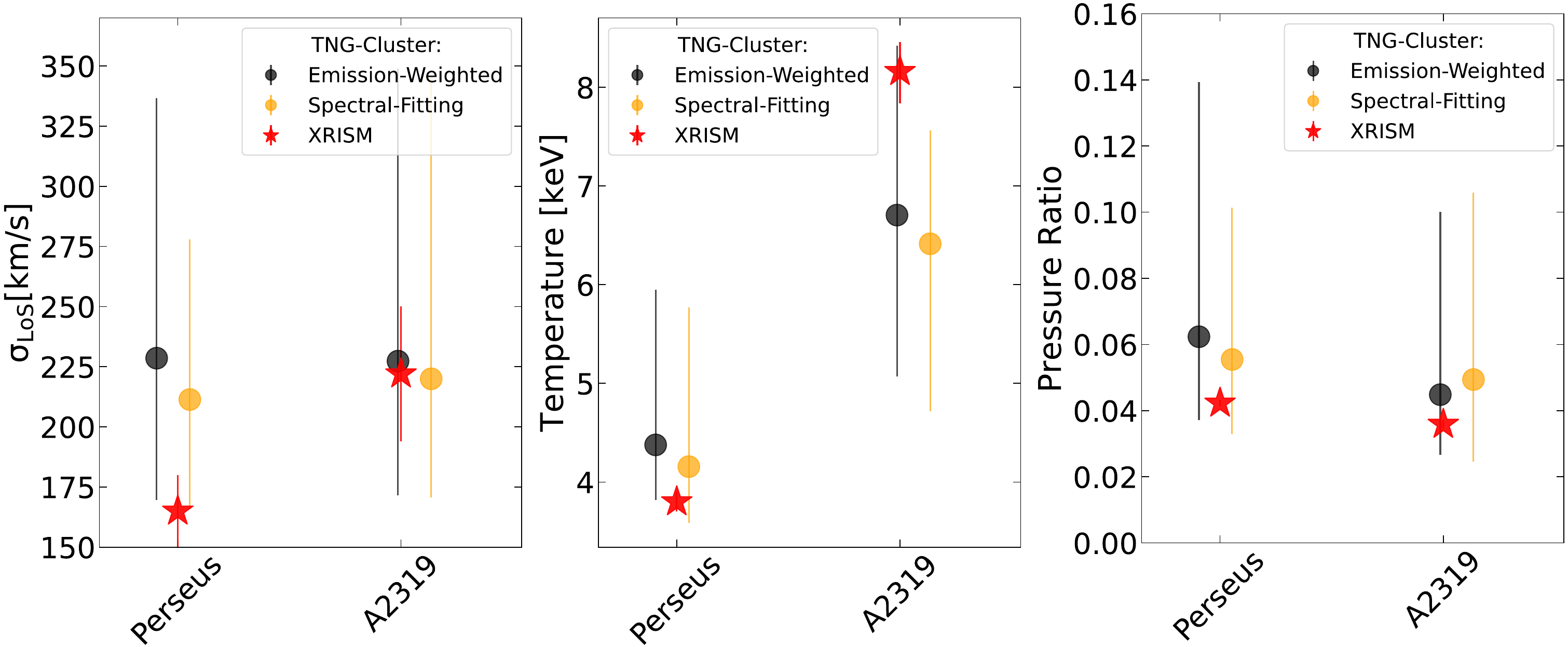}
\caption{Comparison between emission-weighted versus spectral fitting quantities. {\it Top:} X-ray surface brightness map of a Perseus-like cluster in TNG-Cluster (left) and its mock X-ray C0 spectrum (right). {\it Bottom: }Comparisons for two individual systems: Perseus (CC) and A2319 (NCC). The two derivation methods are in close agreement. The XRISM data points are shown for illustration only --- the simulated clusters are not expected to match the observed ones precisely.}
\label{fig:spectral_vs_EW}
\end{figure*}

The {\it top} panels of Figure \ref{fig:spectral_vs_EW} illustrates the mock X-ray analysis for a Perseus-like cluster from the C0 pointing. The {\it top-left} panel shows the intrinsic X-ray map, while the {\it top-right} panel presents the mock spectrum of the C0 pointing, centered on the Fe XXV He$\alpha$ complex, along with the best-fit bapec model. The {\it bottom} panels compare emission-weighted and spectral-fitting results for the two samples of Perseus-like and A2319-like clusters. The two methods yield consistent estimates of temperature and velocity dispersion, as well as the derived kinetic pressure support level. Thus, our adopted emission-weighted method can be reliably used for comparison with observations. 

\bibliography{sample631}{}

\begin{thebibliography}{}
\expandafter\ifx\csname natexlab\endcsname\relax\def\natexlab#1{#1}\fi
\providecommand{\url}[1]{\href{#1}{#1}}
\providecommand{\dodoi}[1]{doi:~\href{http://doi.org/#1}{\nolinkurl{#1}}}
\providecommand{\doeprint}[1]{\href{http://ascl.net/#1}{\nolinkurl{http://ascl.net/#1}}}
\providecommand{\doarXiv}[1]{\href{https://arxiv.org/abs/#1}{\nolinkurl{https://arxiv.org/abs/#1}}}

\bibitem[{{Beck} {et~al.}(2016){Beck}, {Murante}, {Arth}, {Remus}, {Teklu}, {Donnert}, {Planelles}, {Beck}, {F{\"o}rster}, {Imgrund}, {Dolag}, \& {Borgani}}]{beck.etal.2016}
{Beck}, A.~M., {Murante}, G., {Arth}, A., {et~al.} 2016, \mnras, 455, 2110, \dodoi{10.1093/mnras/stv2443}

\bibitem[{{Brunetti} \& {Jones}(2014)}]{brunetti.jones.2014}
{Brunetti}, G., \& {Jones}, T.~W. 2014, International Journal of Modern Physics D, 23, 1430007, \dodoi{10.1142/S0218271814300079}

\bibitem[{{Cavagnolo} {et~al.}(2009){Cavagnolo}, {Donahue}, {Voit}, \& {Sun}}]{cavagnolo.etal.2009}
{Cavagnolo}, K.~W., {Donahue}, M., {Voit}, G.~M., \& {Sun}, M. 2009, \apjs, 182, 12, \dodoi{10.1088/0067-0049/182/1/12}

\bibitem[{{Cui} {et~al.}(2018){Cui}, {Knebe}, {Yepes}, {Pearce}, {Power}, {Dave}, {Arth}, {Borgani}, {Dolag}, {Elahi}, {Mostoghiu}, {Murante}, {Rasia}, {Stoppacher}, {Vega-Ferrero}, {Wang}, {Yang}, {Benson}, {Cora}, {Croton}, {Sinha}, {Stevens}, {Vega-Mart{\'\i}nez}, {Arthur}, {Baldi}, {Ca{\~n}as}, {Cialone}, {Cunnama}, {De Petris}, {Durando}, {Ettori}, {Gottl{\"o}ber}, {Nuza}, {Old}, {Pilipenko}, {Sorce}, \& {Welker}}]{cui.etal.2018}
{Cui}, W., {Knebe}, A., {Yepes}, G., {et~al.} 2018, \mnras, 480, 2898, \dodoi{10.1093/mnras/sty2111}

\bibitem[{{Cui} {et~al.}(2022){Cui}, {Dave}, {Knebe}, {Rasia}, {Gray}, {Pearce}, {Power}, {Yepes}, {Anbajagane}, {Ceverino}, {Contreras-Santos}, {de Andres}, {De Petris}, {Ettori}, {Haggar}, {Li}, {Wang}, {Yang}, {Borgani}, {Dolag}, {Zu}, {Kuchner}, {Ca{\~n}as}, {Ferragamo}, \& {Gianfagna}}]{cui.etal.2022}
{Cui}, W., {Dave}, R., {Knebe}, A., {et~al.} 2022, \mnras, 514, 977, \dodoi{10.1093/mnras/stac1402}

\bibitem[{{Dav{\'e}} {et~al.}(2019){Dav{\'e}}, {Angl{\'e}s-Alc{\'a}zar}, {Narayanan}, {Li}, {Rafieferantsoa}, \& {Appleby}}]{dave.etal.2019}
{Dav{\'e}}, R., {Angl{\'e}s-Alc{\'a}zar}, D., {Narayanan}, D., {et~al.} 2019, \mnras, 486, 2827, \dodoi{10.1093/mnras/stz937}

\bibitem[{{Fabian}(2012)}]{fabian2012}
{Fabian}, A.~C. 2012, \araa, 50, 455, \dodoi{10.1146/annurev-astro-081811-125521}

\bibitem[{{Fujita} {et~al.}(2025){Fujita}, {Fukushima}, {Sato}, {Fukazawa}, \& {Kondo}}]{fujita.etal.2025}
{Fujita}, Y., {Fukushima}, K., {Sato}, K., {Fukazawa}, Y., \& {Kondo}, M. 2025, arXiv e-prints, arXiv:2507.00126, \dodoi{10.48550/arXiv.2507.00126}

\bibitem[{{Fujita} \& {Ohira}(2013)}]{fujita.ohira.2013}
{Fujita}, Y., \& {Ohira}, Y. 2013, \mnras, 428, 599, \dodoi{10.1093/mnras/sts050}

\bibitem[{{George} {et~al.}(2009){George}, {Fabian}, {Sanders}, {Young}, \& {Russell}}]{george.etal.2009}
{George}, M.~R., {Fabian}, A.~C., {Sanders}, J.~S., {Young}, A.~J., \& {Russell}, H.~R. 2009, \mnras, 395, 657, \dodoi{10.1111/j.1365-2966.2009.14547.x}

\bibitem[{{Ghirardini} {et~al.}(2018){Ghirardini}, {Ettori}, {Eckert}, {Molendi}, {Gastaldello}, {Pointecouteau}, {Hurier}, \& {Bourdin}}]{ghirardini.etal.2018}
{Ghirardini}, V., {Ettori}, S., {Eckert}, D., {et~al.} 2018, \aap, 614, A7, \dodoi{10.1051/0004-6361/201731748}

\bibitem[{{Giacintucci} {et~al.}(2020){Giacintucci}, {Markevitch}, {Johnston-Hollitt}, {Wik}, {Wang}, \& {Clarke}}]{giacintucci.etal.2020}
{Giacintucci}, S., {Markevitch}, M., {Johnston-Hollitt}, M., {et~al.} 2020, \apj, 891, 1, \dodoi{10.3847/1538-4357/ab6a9d}

\bibitem[{{Girardi} {et~al.}(2022){Girardi}, {Boschin}, {Nonino}, {Innocentin}, \& {De Grandi}}]{girardi.etal.2022}
{Girardi}, M., {Boschin}, W., {Nonino}, M., {Innocentin}, C., \& {De Grandi}, S. 2022, \aap, 658, A159, \dodoi{10.1051/0004-6361/202142213}

\bibitem[{{Groth} {et~al.}(2025{\natexlab{a}}){Groth}, {Valentini}, {Seidel}, {Vladutescu-Zopp}, {Biffi}, {Dolag}, \& {Sorce}}]{groth.etal.2025b}
{Groth}, F., {Valentini}, M., {Seidel}, B.~A., {et~al.} 2025{\natexlab{a}}, arXiv e-prints, arXiv:2507.02041, \dodoi{10.48550/arXiv.2507.02041}

\bibitem[{{Groth} {et~al.}(2025{\natexlab{b}}){Groth}, {Valentini}, {Steinwandel}, {Vall{\'e}s-P{\'e}rez}, \& {Dolag}}]{groth.etal.2025a}
{Groth}, F., {Valentini}, M., {Steinwandel}, U.~P., {Vall{\'e}s-P{\'e}rez}, D., \& {Dolag}, K. 2025{\natexlab{b}}, \aap, 693, A263, \dodoi{10.1051/0004-6361/202451803}

\bibitem[{{Guo} \& {Oh}(2008)}]{guo.oh.2008}
{Guo}, F., \& {Oh}, S.~P. 2008, \mnras, 384, 251, \dodoi{10.1111/j.1365-2966.2007.12692.x}

\bibitem[{{Heinrich} {et~al.}(2024){Heinrich}, {Zhuravleva}, {Zhang}, {Churazov}, {Forman}, \& {van Weeren}}]{heinrich.etal.2024}
{Heinrich}, A., {Zhuravleva}, I., {Zhang}, C., {et~al.} 2024, \mnras, 528, 7274, \dodoi{10.1093/mnras/stae208}

\bibitem[{{Hitomi Collaboration} {et~al.}(2016){Hitomi Collaboration}, {Aharonian}, {Akamatsu}, {Akimoto}, {Allen}, {Anabuki}, {Angelini}, {Arnaud}, {Audard}, {Awaki}, {Axelsson}, {Bamba}, {Bautz}, {Blandford}, {Brenneman}, {Brown}, {Bulbul}, {Cackett}, {Chernyakova}, {Chiao}, {Coppi}, {Costantini}, {de Plaa}, {den Herder}, {Done}, {Dotani}, {Ebisawa}, {Eckart}, {Enoto}, {Ezoe}, {Fabian}, {Ferrigno}, {Foster}, {Fujimoto}, {Fukazawa}, {Furuzawa}, {Galeazzi}, {Gallo}, {Gandhi}, {Giustini}, {Goldwurm}, {Gu}, {Guainazzi}, {Haba}, {Hagino}, {Hamaguchi}, {Harrus}, {Hatsukade}, {Hayashi}, {Hayashi}, {Hayashida}, {Hiraga}, {Hornschemeier}, {Hoshino}, {Hughes}, {Iizuka}, {Inoue}, {Inoue}, {Ishibashi}, {Ishida}, {Ishikawa}, {Ishisaki}, {Itoh}, {Iyomoto}, {Kaastra}, {Kallman}, {Kamae}, {Kara}, {Kataoka}, {Katsuda}, {Katsuta}, {Kawaharada}, {Kawai}, {Kelley}, {Khangulyan}, {Kilbourne}, {King}, {Kitaguchi}, {Kitamoto}, {Kitayama}, {Kohmura}, {Kokubun}, {Koyama}, {Koyama}, {Kretschmar}, {Krimm}, {Kubota}, {Kunieda},
  {Laurent}, {Lebrun}, {Lee}, {Leutenegger}, {Limousin}, {Loewenstein}, {Long}, {Lumb}, {Madejski}, {Maeda}, {Maier}, {Makishima}, {Markevitch}, {Matsumoto}, {Matsushita}, {McCammon}, {McNamara}, {Mehdipour}, {Miller}, {Miller}, {Mineshige}, {Mitsuda}, {Mitsuishi}, {Miyazawa}, {Mizuno}, {Mori}, {Mori}, {Moseley}, {Mukai}, {Murakami}, {Murakami}, {Mushotzky}, {Nagino}, {Nakagawa}, {Nakajima}, {Nakamori}, {Nakano}, {Nakashima}, {Nakazawa}, {Nobukawa}, {Noda}, {Nomachi}, {O'Dell}, {Odaka}, {Ohashi}, {Ohno}, {Okajima}, {Ota}, {Ozaki}, {Paerels}, {Paltani}, {Parmar}, {Petre}, {Pinto}, {Pohl}, {Porter}, {Pottschmidt}, {Ramsey}, {Reynolds}, {Russell}, {Safi-Harb}, {Saito}, {Sakai}, {Sameshima}, {Sato}, {Sato}, {Sato}, {Sawada}, {Schartel}, {Serlemitsos}, {Seta}, {Shidatsu}, {Simionescu}, {Smith}, {Soong}, {Stawarz}, {Sugawara}, {Sugita}, {Szymkowiak}, {Tajima}, {Takahashi}, {Takahashi}, {Takeda}, {Takei}, {Tamagawa}, {Tamura}, {Tamura}, {Tanaka}, {Tanaka}, {Tanaka}, {Tashiro}, {Tawara}, {Terada}, {Terashima},
  {Tombesi}, {Tomida}, {Tsuboi}, {Tsujimoto}, {Tsunemi}, {Tsuru}, {Uchida}, {Uchiyama}, {Uchiyama}, {Ueda}, {Ueda}, {Ueno}, {Uno}, {Urry}, {Ursino}, {de Vries}, {Watanabe}, {Werner}, {Wik}, {Wilkins}, {Williams}, {Yamada}, {Yamaguchi}, {Yamaoka}, {Yamasaki}, {Yamauchi}, {Yamauchi}, {Yaqoob}, {Yatsu}, {Yonetoku}, {Yoshida}, {Yuasa}, {Zhuravleva}, \& {Zoghbi}}]{hitomi.collaboration.2016}
{Hitomi Collaboration}, {Aharonian}, F., {Akamatsu}, H., {et~al.} 2016, \nat, 535, 117, \dodoi{10.1038/nature18627}

\bibitem[{{Hitomi Collaboration} {et~al.}(2018){Hitomi Collaboration}, {Aharonian}, {Akamatsu}, {Akimoto}, {Allen}, {Angelini}, {Audard}, {Awaki}, {Axelsson}, {Bamba}, {Bautz}, {Blandford}, {Brenneman}, {Brown}, {Bulbul}, {Cackett}, {Canning}, {Chernyakova}, {Chiao}, {Coppi}, {Costantini}, {de Plaa}, {de Vries}, {den Herder}, {Done}, {Dotani}, {Ebisawa}, {Eckart}, {Enoto}, {Ezoe}, {Fabian}, {Ferrigno}, {Foster}, {Fujimoto}, {Fukazawa}, {Furuzawa}, {Galeazzi}, {Gallo}, {Gandhi}, {Giustini}, {Goldwurm}, {Gu}, {Guainazzi}, {Haba}, {Hagino}, {Hamaguchi}, {Harrus}, {Hatsukade}, {Hayashi}, {Hayashi}, {Hayashi}, {Hayashida}, {Hiraga}, {Hornschemeier}, {Hoshino}, {Hughes}, {Ichinohe}, {Iizuka}, {Inoue}, {Inoue}, {Inoue}, {Ishida}, {Ishikawa}, {Ishisaki}, {Iwai}, {Kaastra}, {Kallman}, {Kamae}, {Kataoka}, {Katsuda}, {Kawai}, {Kelley}, {Kilbourne}, {Kitaguchi}, {Kitamoto}, {Kitayama}, {Kohmura}, {Kokubun}, {Koyama}, {Koyama}, {Kretschmar}, {Krimm}, {Kubota}, {Kunieda}, {Laurent}, {Lee}, {Leutenegger}, {Limousin},
  {Loewenstein}, {Long}, {Lumb}, {Madejski}, {Maeda}, {Maier}, {Makishima}, {Markevitch}, {Matsumoto}, {Matsushita}, {McCammon}, {McNamara}, {Mehdipour}, {Miller}, {Miller}, {Mineshige}, {Mitsuda}, {Mitsuishi}, {Miyazawa}, {Mizuno}, {Mori}, {Mori}, {Mukai}, {Murakami}, {Mushotzky}, {Nakagawa}, {Nakajima}, {Nakamori}, {Nakashima}, {Nakazawa}, {Nobukawa}, {Nobukawa}, {Noda}, {Odaka}, {Ohashi}, {Ohno}, {Okajima}, {Ota}, {Ozaki}, {Paerels}, {Paltani}, {Petre}, {Pinto}, {Porter}, {Pottschmidt}, {Reynolds}, {Safi-Harb}, {Saito}, {Sakai}, {Sasaki}, {Sato}, {Sato}, {Sato}, {Sawada}, {Schartel}, {Serlemtsos}, {Seta}, {Shidatsu}, {Simionescu}, {Smith}, {Soong}, {Stawarz}, {Sugawara}, {Sugita}, {Szymkowiak}, {Tajima}, {Takahashi}, {Takahashi}, {Takeda}, {Takei}, {Tamagawa}, {Tamura}, {Tanaka}, {Tanaka}, {Tanaka}, {Tanaka}, {Tashiro}, {Tawara}, {Terada}, {Terashima}, {Tombesi}, {Tomida}, {Tsuboi}, {Tsujimoto}, {Tsunemi}, {Tsuru}, {Uchida}, {Uchiyama}, {Uchiyama}, {Ueda}, {Ueda}, {Uno}, {Urry}, {Ursino}, {Wang},
  {Watanabe}, {Werner}, {Wilkins}, {Williams}, {Yamada}, {Yamaguchi}, {Yamaoka}, {Yamasaki}, {Yamauchi}, {Yamauchi}, {Yaqoob}, {Yatsu}, {Yonetoku}, {Zhuravleva}, \& {Zoghbi}}]{hitomi.2018}
---. 2018, \pasj, 70, 9, \dodoi{10.1093/pasj/psx138}

\bibitem[{{Ho} {et~al.}(2022){Ho}, {Ntampaka}, {Rau}, {Chen}, {Lansberry}, {Ruehle}, \& {Trac}}]{ho.etal.2022}
{Ho}, M., {Ntampaka}, M., {Rau}, M.~M., {et~al.} 2022, Nature Astronomy, 6, 936, \dodoi{10.1038/s41550-022-01711-1}

\bibitem[{{Hopkins}(2015)}]{hopkins.etal.2015}
{Hopkins}, P.~F. 2015, \mnras, 450, 53, \dodoi{10.1093/mnras/stv195}

\bibitem[{{Jacob} \& {Pfrommer}(2017)}]{jacob.pfrommer.2017}
{Jacob}, S., \& {Pfrommer}, C. 2017, \mnras, 467, 1478, \dodoi{10.1093/mnras/stx132}

\bibitem[{{Klypin} {et~al.}(2016){Klypin}, {Yepes}, {Gottl{\"o}ber}, {Prada}, \& {He{\ss}}}]{klypin.etal.2016}
{Klypin}, A., {Yepes}, G., {Gottl{\"o}ber}, S., {Prada}, F., \& {He{\ss}}, S. 2016, \mnras, 457, 4340, \dodoi{10.1093/mnras/stw248}

\bibitem[{{Kravtsov} \& {Borgani}(2012)}]{kravtsov.borgani.2012}
{Kravtsov}, A.~V., \& {Borgani}, S. 2012, \araa, 50, 353, \dodoi{10.1146/annurev-astro-081811-125502}

\bibitem[{{Kunz} {et~al.}(2022){Kunz}, {Jones}, \& {Zhuravleva}}]{kunz.etal.2022}
{Kunz}, M.~W., {Jones}, T.~W., \& {Zhuravleva}, I. 2022, in Handbook of X-ray and Gamma-ray Astrophysics, ed. C.~{Bambi} \& A.~{Sangangelo}, 56, \dodoi{10.1007/978-981-16-4544-0_125-1}

\bibitem[{{Lehle} {et~al.}(2024){Lehle}, {Nelson}, {Pillepich}, {Truong}, \& {Rohr}}]{lehle.etal.2024}
{Lehle}, K., {Nelson}, D., {Pillepich}, A., {Truong}, N., \& {Rohr}, E. 2024, \aap, 687, A129, \dodoi{10.1051/0004-6361/202348609}

\bibitem[{{Lieu} {et~al.}(2016){Lieu}, {Smith}, {Giles}, {Ziparo}, {Maughan}, {D{\'e}mocl{\`e}s}, {Pacaud}, {Pierre}, {Adami}, {Bah{\'e}}, {Clerc}, {Chiappetti}, {Eckert}, {Ettori}, {Lavoie}, {Le Fevre}, {McCarthy}, {Kilbinger}, {Ponman}, {Sadibekova}, \& {Willis}}]{lieu.etal.2016}
{Lieu}, M., {Smith}, G.~P., {Giles}, P.~A., {et~al.} 2016, \aap, 592, A4, \dodoi{10.1051/0004-6361/201526883}

\bibitem[{{Mahdavi} {et~al.}(2013){Mahdavi}, {Hoekstra}, {Babul}, {Bildfell}, {Jeltema}, \& {Henry}}]{mahdavi.etal.2013}
{Mahdavi}, A., {Hoekstra}, H., {Babul}, A., {et~al.} 2013, \apj, 767, 116, \dodoi{10.1088/0004-637X/767/2/116}

\bibitem[{{Marinacci} {et~al.}(2018){Marinacci}, {Vogelsberger}, {Pakmor}, {Torrey}, {Springel}, {Hernquist}, {Nelson}, {Weinberger}, {Pillepich}, {Naiman}, \& {Genel}}]{marinacci.etal.2018}
{Marinacci}, F., {Vogelsberger}, M., {Pakmor}, R., {et~al.} 2018, \mnras, 480, 5113, \dodoi{10.1093/mnras/sty2206}

\bibitem[{{McDonald} {et~al.}(2013){McDonald}, {Benson}, {Vikhlinin}, {Stalder}, {Bleem}, {de Haan}, {Lin}, {Aird}, {Ashby}, {Bautz}, {Bayliss}, {Bocquet}, {Brodwin}, {Carlstrom}, {Chang}, {Cho}, {Clocchiatti}, {Crawford}, {Crites}, {Desai}, {Dobbs}, {Dudley}, {Foley}, {Forman}, {George}, {Gettings}, {Gladders}, {Gonzalez}, {Halverson}, {High}, {Holder}, {Holzapfel}, {Hoover}, {Hrubes}, {Jones}, {Joy}, {Keisler}, {Knox}, {Lee}, {Leitch}, {Liu}, {Lueker}, {Luong-Van}, {Mantz}, {Marrone}, {McMahon}, {Mehl}, {Meyer}, {Miller}, {Mocanu}, {Mohr}, {Montroy}, {Murray}, {Nurgaliev}, {Padin}, {Plagge}, {Pryke}, {Reichardt}, {Rest}, {Ruel}, {Ruhl}, {Saliwanchik}, {Saro}, {Sayre}, {Schaffer}, {Shirokoff}, {Song}, {{\v{S}}uhada}, {Spieler}, {Stanford}, {Staniszewski}, {Stark}, {Story}, {van Engelen}, {Vanderlinde}, {Vieira}, {Williamson}, {Zahn}, \& {Zenteno}}]{mcdonald.etal.2013}
{McDonald}, M., {Benson}, B.~A., {Vikhlinin}, A., {et~al.} 2013, \apj, 774, 23, \dodoi{10.1088/0004-637X/774/1/23}

\bibitem[{{Naiman} {et~al.}(2018){Naiman}, {Pillepich}, {Springel}, {Ramirez-Ruiz}, {Torrey}, {Vogelsberger}, {Pakmor}, {Nelson}, {Marinacci}, {Hernquist}, {Weinberger}, \& {Genel}}]{naiman.etal.2018}
{Naiman}, J.~P., {Pillepich}, A., {Springel}, V., {et~al.} 2018, \mnras, 477, 1206, \dodoi{10.1093/mnras/sty618}

\bibitem[{{Nelson} {et~al.}(2024){Nelson}, {Pillepich}, {Ayromlou}, {Lee}, {Lehle}, {Rohr}, \& {Truong}}]{nelson.etal.2024}
{Nelson}, D., {Pillepich}, A., {Ayromlou}, M., {et~al.} 2024, \aap, 686, A157, \dodoi{10.1051/0004-6361/202348608}

\bibitem[{{Nelson} {et~al.}(2018){Nelson}, {Pillepich}, {Springel}, {Weinberger}, {Hernquist}, {Pakmor}, {Genel}, {Torrey}, {Vogelsberger}, {Kauffmann}, {Marinacci}, \& {Naiman}}]{nelson.etal.2018}
{Nelson}, D., {Pillepich}, A., {Springel}, V., {et~al.} 2018, \mnras, 475, 624, \dodoi{10.1093/mnras/stx3040}

\bibitem[{{Nelson} {et~al.}(2019{\natexlab{a}}){Nelson}, {Springel}, {Pillepich}, {Rodriguez-Gomez}, {Torrey}, {Genel}, {Vogelsberger}, {Pakmor}, {Marinacci}, {Weinberger}, {Kelley}, {Lovell}, {Diemer}, \& {Hernquist}}]{nelson.etal.2019a}
{Nelson}, D., {Springel}, V., {Pillepich}, A., {et~al.} 2019{\natexlab{a}}, Computational Astrophysics and Cosmology, 6, 2, \dodoi{10.1186/s40668-019-0028-x}

\bibitem[{{Nelson} {et~al.}(2019{\natexlab{b}}){Nelson}, {Pillepich}, {Springel}, {Pakmor}, {Weinberger}, {Genel}, {Torrey}, {Vogelsberger}, {Marinacci}, \& {Hernquist}}]{nelson.etal.2019b}
{Nelson}, D., {Pillepich}, A., {Springel}, V., {et~al.} 2019{\natexlab{b}}, \mnras, 490, 3234, \dodoi{10.1093/mnras/stz2306}

\bibitem[{{Peres} {et~al.}(1998){Peres}, {Fabian}, {Edge}, {Allen}, {Johnstone}, \& {White}}]{peres.etal.1998}
{Peres}, C.~B., {Fabian}, A.~C., {Edge}, A.~C., {et~al.} 1998, \mnras, 298, 416, \dodoi{10.1046/j.1365-8711.1998.01624.x}

\bibitem[{{Pillepich} {et~al.}(2018){Pillepich}, {Nelson}, {Hernquist}, {Springel}, {Pakmor}, {Torrey}, {Weinberger}, {Genel}, {Naiman}, {Marinacci}, \& {Vogelsberger}}]{pillepich.etal.2018b}
{Pillepich}, A., {Nelson}, D., {Hernquist}, L., {et~al.} 2018, \mnras, 475, 648, \dodoi{10.1093/mnras/stx3112}

\bibitem[{{Planck Collaboration} {et~al.}(2016){Planck Collaboration}, {Ade}, {Aghanim}, {Arnaud}, {Ashdown}, {Aumont}, {Baccigalupi}, {Banday}, {Barreiro}, {Bartlett}, {Bartolo}, {Battaner}, {Battye}, {Benabed}, {Beno{\^\i}t}, {Benoit-L{\'e}vy}, {Bernard}, {Bersanelli}, {Bielewicz}, {Bock}, {Bonaldi}, {Bonavera}, {Bond}, {Borrill}, {Bouchet}, {Boulanger}, {Bucher}, {Burigana}, {Butler}, {Calabrese}, {Cardoso}, {Catalano}, {Challinor}, {Chamballu}, {Chary}, {Chiang}, {Chluba}, {Christensen}, {Church}, {Clements}, {Colombi}, {Colombo}, {Combet}, {Coulais}, {Crill}, {Curto}, {Cuttaia}, {Danese}, {Davies}, {Davis}, {de Bernardis}, {de Rosa}, {de Zotti}, {Delabrouille}, {D{\'e}sert}, {Di Valentino}, {Dickinson}, {Diego}, {Dolag}, {Dole}, {Donzelli}, {Dor{\'e}}, {Douspis}, {Ducout}, {Dunkley}, {Dupac}, {Efstathiou}, {Elsner}, {En{\ss}lin}, {Eriksen}, {Farhang}, {Fergusson}, {Finelli}, {Forni}, {Frailis}, {Fraisse}, {Franceschi}, {Frejsel}, {Galeotta}, {Galli}, {Ganga}, {Gauthier}, {Gerbino}, {Ghosh}, {Giard},
  {Giraud-H{\'e}raud}, {Giusarma}, {Gjerl{\o}w}, {Gonz{\'a}lez-Nuevo}, {G{\'o}rski}, {Gratton}, {Gregorio}, {Gruppuso}, {Gudmundsson}, {Hamann}, {Hansen}, {Hanson}, {Harrison}, {Helou}, {Henrot-Versill{\'e}}, {Hern{\'a}ndez-Monteagudo}, {Herranz}, {Hildebrandt}, {Hivon}, {Hobson}, {Holmes}, {Hornstrup}, {Hovest}, {Huang}, {Huffenberger}, {Hurier}, {Jaffe}, {Jaffe}, {Jones}, {Juvela}, {Keih{\"a}nen}, {Keskitalo}, {Kisner}, {Kneissl}, {Knoche}, {Knox}, {Kunz}, {Kurki-Suonio}, {Lagache}, {L{\"a}hteenm{\"a}ki}, {Lamarre}, {Lasenby}, {Lattanzi}, {Lawrence}, {Leahy}, {Leonardi}, {Lesgourgues}, {Levrier}, {Lewis}, {Liguori}, {Lilje}, {Linden-V{\o}rnle}, {L{\'o}pez-Caniego}, {Lubin}, {Mac{\'\i}as-P{\'e}rez}, {Maggio}, {Maino}, {Mandolesi}, {Mangilli}, {Marchini}, {Maris}, {Martin}, {Martinelli}, {Mart{\'\i}nez-Gonz{\'a}lez}, {Masi}, {Matarrese}, {McGehee}, {Meinhold}, {Melchiorri}, {Melin}, {Mendes}, {Mennella}, {Migliaccio}, {Millea}, {Mitra}, {Miville-Desch{\^e}nes}, {Moneti}, {Montier}, {Morgante}, {Mortlock},
  {Moss}, {Munshi}, {Murphy}, {Naselsky}, {Nati}, {Natoli}, {Netterfield}, {N{\o}rgaard-Nielsen}, {Noviello}, {Novikov}, {Novikov}, {Oxborrow}, {Paci}, {Pagano}, {Pajot}, {Paladini}, {Paoletti}, {Partridge}, {Pasian}, {Patanchon}, {Pearson}, {Perdereau}, {Perotto}, {Perrotta}, {Pettorino}, {Piacentini}, {Piat}, {Pierpaoli}, {Pietrobon}, {Plaszczynski}, {Pointecouteau}, {Polenta}, {Popa}, {Pratt}, {Pr{\'e}zeau}, {Prunet}, {Puget}, {Rachen}, {Reach}, {Rebolo}, {Reinecke}, {Remazeilles}, {Renault}, {Renzi}, {Ristorcelli}, {Rocha}, {Rosset}, {Rossetti}, {Roudier}, {Rouill{\'e} d'Orfeuil}, {Rowan-Robinson}, {Rubi{\~n}o-Mart{\'\i}n}, {Rusholme}, {Said}, {Salvatelli}, {Salvati}, {Sandri}, {Santos}, {Savelainen}, {Savini}, {Scott}, {Seiffert}, {Serra}, {Shellard}, {Spencer}, {Spinelli}, {Stolyarov}, {Stompor}, {Sudiwala}, {Sunyaev}, {Sutton}, {Suur-Uski}, {Sygnet}, {Tauber}, {Terenzi}, {Toffolatti}, {Tomasi}, {Tristram}, {Trombetti}, {Tucci}, {Tuovinen}, {T{\"u}rler}, {Umana}, {Valenziano}, {Valiviita}, {Van Tent},
  {Vielva}, {Villa}, {Wade}, {Wandelt}, {Wehus}, {White}, {White}, {Wilkinson}, {Yvon}, {Zacchei}, \& {Zonca}}]{planck.collaboration.2016}
{Planck Collaboration}, {Ade}, P.~A.~R., {Aghanim}, N., {et~al.} 2016, \aap, 594, A13, \dodoi{10.1051/0004-6361/201525830}

\bibitem[{{Rasia} {et~al.}(2015){Rasia}, {Borgani}, {Murante}, {Planelles}, {Beck}, {Biffi}, {Ragone-Figueroa}, {Granato}, {Steinborn}, \& {Dolag}}]{rasia.etal.2015}
{Rasia}, E., {Borgani}, S., {Murante}, G., {et~al.} 2015, \apjl, 813, L17, \dodoi{10.1088/2041-8205/813/1/L17}

\bibitem[{{Reiprich} \& {B{\"o}hringer}(2002)}]{reiprich.bohringer.2002}
{Reiprich}, T.~H., \& {B{\"o}hringer}, H. 2002, \apj, 567, 716, \dodoi{10.1086/338753}

\bibitem[{{Rose} {et~al.}(2025){Rose}, {McNamara}, {Meunier}, {Fabian}, {Russell}, {Nulsen}, {Dizdar}, {Heckman}, {McDonald}, {Markevitch}, {Paerels}, {Simionescu}, {Werner}, {Coil}, {Hodges-Kluck}, {Miller}, \& {Wise}}]{rose.etal.2025}
{Rose}, T., {McNamara}, B.~R., {Meunier}, J., {et~al.} 2025, arXiv e-prints, arXiv:2505.01494, \dodoi{10.48550/arXiv.2505.01494}

\bibitem[{{Sanders} {et~al.}(2020){Sanders}, {Dennerl}, {Russell}, {Eckert}, {Pinto}, {Fabian}, {Walker}, {Tamura}, {ZuHone}, \& {Hofmann}}]{sanders.etal.2020}
{Sanders}, J.~S., {Dennerl}, K., {Russell}, H.~R., {et~al.} 2020, \aap, 633, A42, \dodoi{10.1051/0004-6361/201936468}

\bibitem[{{Sarkar} {et~al.}(2025){Sarkar}, {Miller}, {Ota}, {Kilbourne}, {McNamara}, {Sun}, {Lovisari}, {Ettori}, {Eckert}, {Szymkowiak}, {Bartalesi}, \& {Loewenstein}}]{xrism.a2029.2025c}
{Sarkar}, A., {Miller}, E., {Ota}, N., {et~al.} 2025, \pasj, 77, S254, \dodoi{10.1093/pasj/psaf093}

\bibitem[{{Sayers} {et~al.}(2021){Sayers}, {Sereno}, {Ettori}, {Rasia}, {Cui}, {Golwala}, {Umetsu}, \& {Yepes}}]{sayers.etal.2021}
{Sayers}, J., {Sereno}, M., {Ettori}, S., {et~al.} 2021, \mnras, 505, 4338, \dodoi{10.1093/mnras/stab1542}

\bibitem[{{Schellenberger} {et~al.}(2015){Schellenberger}, {Reiprich}, {Lovisari}, {Nevalainen}, \& {David}}]{schellenberger.etal.2015}
{Schellenberger}, G., {Reiprich}, T.~H., {Lovisari}, L., {Nevalainen}, J., \& {David}, L. 2015, \aap, 575, A30, \dodoi{10.1051/0004-6361/201424085}

\bibitem[{{Simionescu} {et~al.}(2017){Simionescu}, {Werner}, {Mantz}, {Allen}, \& {Urban}}]{simionescu.etal.2017}
{Simionescu}, A., {Werner}, N., {Mantz}, A., {Allen}, S.~W., \& {Urban}, O. 2017, \mnras, 469, 1476, \dodoi{10.1093/mnras/stx919}

\bibitem[{{Simionescu} {et~al.}(2011){Simionescu}, {Allen}, {Mantz}, {Werner}, {Takei}, {Morris}, {Fabian}, {Sanders}, {Nulsen}, {George}, \& {Taylor}}]{simionescu.etal.2011}
{Simionescu}, A., {Allen}, S.~W., {Mantz}, A., {et~al.} 2011, Science, 331, 1576, \dodoi{10.1126/science.1200331}

\bibitem[{{Smith} {et~al.}(2001){Smith}, {Brickhouse}, {Liedahl}, \& {Raymond}}]{smith.etal.2001}
{Smith}, R.~K., {Brickhouse}, N.~S., {Liedahl}, D.~A., \& {Raymond}, J.~C. 2001, \apjl, 556, L91, \dodoi{10.1086/322992}

\bibitem[{{Springel} {et~al.}(2018){Springel}, {Pakmor}, {Pillepich}, {Weinberger}, {Nelson}, {Hernquist}, {Vogelsberger}, {Genel}, {Torrey}, {Marinacci}, \& {Naiman}}]{springel.etal.2018}
{Springel}, V., {Pakmor}, R., {Pillepich}, A., {et~al.} 2018, \mnras, 475, 676, \dodoi{10.1093/mnras/stx3304}

\bibitem[{{Truong} {et~al.}(2024){Truong}, {Pillepich}, {Nelson}, {Zhuravleva}, {Lee}, {Ayromlou}, \& {Lehle}}]{truong.etal.2024}
{Truong}, N., {Pillepich}, A., {Nelson}, D., {et~al.} 2024, \aap, 686, A200, \dodoi{10.1051/0004-6361/202348562}

\bibitem[{{Truong} {et~al.}(2018){Truong}, {Rasia}, {Mazzotta}, {Planelles}, {Biffi}, {Fabjan}, {Beck}, {Borgani}, {Dolag}, {Gaspari}, {Granato}, {Murante}, {Ragone-Figueroa}, \& {Steinborn}}]{truong.etal.2018}
{Truong}, N., {Rasia}, E., {Mazzotta}, P., {et~al.} 2018, \mnras, 474, 4089, \dodoi{10.1093/mnras/stx2927}

\bibitem[{{Vazza} \& {Brunetti}(2025)}]{vazza.brunetti.2025}
{Vazza}, F., \& {Brunetti}, G. 2025, arXiv e-prints, arXiv:2507.04727, \dodoi{10.48550/arXiv.2507.04727}

\bibitem[{{Walker} {et~al.}(2012){Walker}, {Fabian}, {Sanders}, {George}, \& {Tawara}}]{walker.etal.2012}
{Walker}, S.~A., {Fabian}, A.~C., {Sanders}, J.~S., {George}, M.~R., \& {Tawara}, Y. 2012, \mnras, 422, 3503, \dodoi{10.1111/j.1365-2966.2012.20860.x}

\bibitem[{{Weinberger} {et~al.}(2017){Weinberger}, {Springel}, {Hernquist}, {Pillepich}, {Marinacci}, {Pakmor}, {Nelson}, {Genel}, {Vogelsberger}, {Naiman}, \& {Torrey}}]{weinberger.etal.2017}
{Weinberger}, R., {Springel}, V., {Hernquist}, L., {et~al.} 2017, \mnras, 465, 3291, \dodoi{10.1093/mnras/stw2944}

\bibitem[{{XRISM Collaboration} {et~al.}(2025{\natexlab{a}}){XRISM Collaboration}, {Audard}, {Awaki}, {Ballhausen}, {Bamba}, {Behar}, {Boissay-Malaquin}, {Brenneman}, {Brown}, {Corrales}, {Costantini}, {Cumbee}, {Done}, {Dotani}, {Ebisawa}, {Eckart}, {Eckert}, {Enoto}, {Eguchi}, {Ezoe}, {Foster}, {Fujimoto}, {Fujita}, {Fukazawa}, {Fukushima}, {Furuzawa}, {Gallo}, {Garc{\'\i}a}, {Gu}, {Guainazzi}, {Hagino}, {Hamaguchi}, {Hatsukade}, {Hayashi}, {Hayashi}, {Hell}, {Hodges-Kluck}, {Hornschemeier}, {Ichinohe}, {Ishida}, {Ishikawa}, {Ishisaki}, {Kaastra}, {Kallman}, {Kara}, {Katsuda}, {Kanemaru}, {Kelley}, {Kilbourne}, {Kitamoto}, {Kobayashi}, {Kohmura}, {Kubota}, {Leutenegger}, {Loewenstein}, {Maeda}, {Markevitch}, {Matsumoto}, {Matsushita}, {McCammon}, {McNamara}, {Mernier}, {Miller}, {Miller}, {Mitsuishi}, {Mizumoto}, {Mizuno}, {Mori}, {Mukai}, {Murakami}, {Mushotzky}, {Nakajima}, {Nakazawa}, {Ness}, {Nobukawa}, {Nobukawa}, {Noda}, {Odaka}, {Ogawa}, {Ogorzalek}, {Okajima}, {Ota}, {Paltani}, {Petre}, {Plucinsky},
  {Porter}, {Pottschmidt}, {Sato}, {Sato}, {Sawada}, {Seta}, {Shidatsu}, {Simionescu}, {Smith}, {Suzuki}, {Szymkowiak}, {Takahashi}, {Takeo}, {Tamagawa}, {Tamura}, {Tanaka}, {Tanimoto}, {Tashiro}, {Terada}, {Terashima}, {Trigo}, {Tsuboi}, {Tsujimoto}, {Tsunemi}, {Tsuru}, {Uchida}, {Uchida}, {Uchida}, {Uchiyama}, {Ueda}, {Uno}, {Vink}, {Watanabe}, {Williams}, {Yamada}, {Yamada}, {Yamaguchi}, {Yamaoka}, {Yamasaki}, {Yamauchi}, {Yamauchi}, {Yaqoob}, {Yoneyama}, {Yoshida}, {Yukita}, {Zhuravleva}, {Kondo}, {Werner}, {Pl{\v{s}}ek}, {Sun}, {Hosogi}, \& {Majumder}}]{xrism.centaurus.2025}
{XRISM Collaboration}, {Audard}, M., {Awaki}, H., {et~al.} 2025{\natexlab{a}}, \nat, 638, 365, \dodoi{10.1038/s41586-024-08561-z}

\bibitem[{{XRISM Collaboration} {et~al.}(2025{\natexlab{b}}){XRISM Collaboration}, {Audard}, {Awaki}, {Ballhausen}, {Bamba}, {Behar}, {Boissay-Malaquin}, {Brenneman}, {Brown}, {Corrales}, {Costantini}, {Cumbee}, {Diaz Trigo}, {Done}, {Dotani}, {Ebisawa}, {Eckart}, {Eckert}, {Eguchi}, {Enoto}, {Ezoe}, {Foster}, {Fujimoto}, {Fujita}, {Fukazawa}, {Fukushima}, {Furuzawa}, {Gallo}, {Garcia}, {Gu}, {Guainazzi}, {Hagino}, {Hamaguchi}, {Hatsukade}, {Hayashi}, {Hayashi}, {Hell}, {Hodges-Kluck}, {Hornschemeier}, {Ichinohe}, {Ishi}, {Ishida}, {Ishikawa}, {Ishisaki}, {Kaastra}, {Kallman}, {Kara}, {Katsuda}, {Kanemaru}, {Kelley}, {Kilbourne}, {Kitamoto}, {Kobayashi}, {Kohmura}, {Kubota}, {Leutenegger}, {Loewenstein}, {Maeda}, {Markevitch}, {Matsumoto}, {Matsushita}, {McCammon}, {McNamara}, {Mernier}, {Miller}, {Miller}, {Mitsuishi}, {Mizumoto}, {Mizuno}, {Mori}, {Mukai}, {Murakami}, {Mushotzky}, {Nakajima}, {Nakazawa}, {Ness}, {Nobukawa}, {Nobukawa}, {Noda}, {Odaka}, {Ogawa}, {Ogorzalek}, {Okajima}, {Ota}, {Paltani},
  {Petre}, {Plucinsky}, {Porter}, {Pottschmidt}, {Sato}, {Sato}, {Sawada}, {Seta}, {Shidatsu}, {Simionescu}, {Smith}, {Suzuki}, {Szymkowiak}, {Takahashi}, {Takeo}, {Tamagawa}, {Tamura}, {Tanaka}, {Tanimoto}, {Tashiro}, {Terada}, {Terashima}, {Tsuboi}, {Tsujimoto}, {Tsunemi}, {Tsuru}, {Tumer}, {Uchida}, {Uchida}, {Uchida}, {Uchiyama}, {Ueda}, {Uno}, {Vink}, {Watanabe}, {Williams}, {Yamada}, {Yamada}, {Yamaguchi}, {Yamaoka}, {Yamasaki}, {Yamauchi}, {Yamauchi}, {Yaqoob}, {Yoneyama}, {Yoshida}, {Yukita}, {Zhuravleva}, {Bellomi}, {Drury}, {Heinrich}, {Hlavacek-Larrondo}, {Meunier}, {Migkas}, {Shefler}, {Stancil}, {Truong}, {Ueda}, {Vigneron}, {Zhang}, \& {ZuHone}}]{xrism.perseus.2025}
---. 2025{\natexlab{b}}, arXiv e-prints, arXiv:2509.04421, \dodoi{10.48550/arXiv.2509.04421}

\bibitem[{{Xrism Collaboration} {et~al.}(2025){Xrism Collaboration}, {Audard}, {Awaki}, {Ballhausen}, {Bamba}, {Behar}, {Boissay-Malaquin}, {Brenneman}, {Brown}, {Corrales}, {Costantini}, {Cumbee}, {Diaz Trigo}, {Done}, {Dotani}, {Ebisawa}, {Eckart}, {Eckert}, {Eguchi}, {Enoto}, {Ezoe}, {Foster}, {Fujimoto}, {Fujita}, {Fukazawa}, {Fukushima}, {Furuzawa}, {Gallo}, {Garc{\'\i}a}, {Gu}, {Guainazzi}, {Hagino}, {Hamaguchi}, {Hatsukade}, {Hayashi}, {Hayashi}, {Hell}, {Hodges-Kluck}, {Hornschemeier}, {Ichinohe}, {Ishida}, {Ishikawa}, {Ishisaki}, {Kaastra}, {Kallman}, {Kara}, {Katsuda}, {Kanemaru}, {Kelley}, {Kilbourne}, {Kitamoto}, {Kobayashi}, {Kohmura}, {Kubota}, {Leutenegger}, {Loewenstein}, {Maeda}, {Markevitch}, {Matsumoto}, {Matsushita}, {McCammon}, {McNamara}, {Mernier}, {Miller}, {Miller}, {Mitsuishi}, {Mizumoto}, {Mizuno}, {Mori}, {Mukai}, {Murakami}, {Mushotzky}, {Nakajima}, {Nakazawa}, {Ness}, {Nobukawa}, {Nobukawa}, {Noda}, {Odaka}, {Ogawa}, {Ogorzalek}, {Okajima}, {Ota}, {Paltani}, {Petre}, {Plucinsky},
  {Porter}, {Pottschmidt}, {Sato}, {Sato}, {Sawada}, {Seta}, {Shidatsu}, {Simionescu}, {Smith}, {Suzuki}, {Szymkowiak}, {Takahashi}, {Takeo}, {Tamagawa}, {Tamura}, {Tanaka}, {Tanimoto}, {Tashiro}, {Terada}, {Terashima}, {Tsuboi}, {Tsujimoto}, {Tsunemi}, {Tsuru}, {Uchida}, {Uchida}, {Uchida}, {Uchiyama}, {Ueda}, {Uno}, {Vink}, {Watanabe}, {Williams}, {Yamada}, {Yamada}, {Yamaguchi}, {Yamaoka}, {Yamasaki}, {Yamauchi}, {Yamauchi}, {Yaqoob}, {Yoneyama}, {Yoshida}, {Yukita}, {Zhuravleva}, {Bartalesi}, {Ettori}, {Kosarzycki}, {Lovisari}, {Rose}, {Sarkar}, {Sun}, \& {Tamhane}}]{xrism.a2029.2025a}
{Xrism Collaboration}, {Audard}, M., {Awaki}, H., {et~al.} 2025, \apjl, 982, L5, \dodoi{10.3847/2041-8213/ada7cd}

\bibitem[{{XRISM Collaboration} {et~al.}(2025{\natexlab{a}}){XRISM Collaboration}, {Audard}, {Awaki}, {Ballhausen}, {Bamba}, {Behar}, {Boissay-Malaquin}, {Brenneman}, {Brown}, {Corrales}, {Costantini}, {Cumbee}, {Diaz Trigo}, {Done}, {Dotani}, {Ebisawa}, {Eckart}, {Eckert}, {Eguchi}, {Enoto}, {Ezoe}, {Foster}, {Fujimoto}, {Fujita}, {Fukazawa}, {Fukushima}, {Furuzawa}, {Gallo}, {Garc{\'\i}a}, {Gu}, {Guainazzi}, {Hagino}, {Hamaguchi}, {Hatsukade}, {Hayashi}, {Hayashi}, {Hell}, {Hodges-Kluck}, {Hornschemeier}, {Ichinohe}, {Ishi}, {Ishida}, {Ishikawa}, {Ishisaki}, {Kaastra}, {Kallman}, {Kara}, {Katsuda}, {Kanemaru}, {Kelley}, {Kilbourne}, {Kitamoto}, {Kobayashi}, {Kohmura}, {Kubota}, {Leutenegger}, {Loewenstein}, {Maeda}, {Markevitch}, {Matsumoto}, {Matsushita}, {McCammon}, {McNamara}, {Mernier}, {Miller}, {Miller}, {Mitsuishi}, {Mizumoto}, {Mizuno}, {Mori}, {Mukai}, {Murakami}, {Mushotzky}, {Nakajima}, {Nakazawa}, {Ness}, {Nobukawa}, {Nobukawa}, {Noda}, {Odaka}, {Ogawa}, {Ogorzalek}, {Okajima}, {Ota}, {Paltani},
  {Petre}, {Plucinsky}, {Porter}, {Pottschmidt}, {Sato}, {Sato}, {Sawada}, {Seta}, {Shidatsu}, {Simionescu}, {Smith}, {Suzuki}, {Szymkowiak}, {Takahashi}, {Takeo}, {Tamagawa}, {Tamura}, {Tanaka}, {Tanimoto}, {Tashiro}, {Terada}, {Terashima}, {Tsuboi}, {Tsujimoto}, {Tsunemi}, {Tsuru}, {Uchida}, {Uchida}, {Uchida}, {Uchiyama}, {Ueda}, {Uno}, {Vink}, {Watanabe}, {Williams}, {Yamada}, {Yamada}, {Yamaguchi}, {Yamaoka}, {Yamasaki}, {Yamauchi}, {Yamauchi}, {Yaqoob}, {Yoneyama}, {Yoshida}, {Yukita}, {Zhuravleva}, {Bartalesi}, {Ettori}, {Kosarzycki}, {Lovisari}, {Rose}, {Sarkar}, {Sun}, \& {Tamhane}}]{xrism.a2029.2025b}
{XRISM Collaboration}, {Audard}, M., {Awaki}, H., {et~al.} 2025{\natexlab{a}}, arXiv e-prints, arXiv:2505.06533, \dodoi{10.48550/arXiv.2505.06533}

\bibitem[{{XRISM Collaboration} {et~al.}(2025{\natexlab{b}}){XRISM Collaboration}, {Audard}, {Awaki}, {Ballhausen}, {Bamba}, {Behar}, {Boissay-malaquin}, {Brenneman}, {Brown}, {Corrales}, {Costantini}, {Cumbee}, {Diaz Trigo}, {Done}, {Dotani}, {Ebisawa}, {Eckart}, {Eckert}, {Eguchi}, {Enoto}, {Ezoe}, {Foster}, {Fujimoto}, {Fujita}, {Fukazawa}, {Fukushima}, {Furuzawa}, {Gallo}, {Garc{\'\i}a}, {Gu}, {Guainazzi}, {Hagino}, {Hamaguchi}, {Hatsukade}, {Hayashi}, {Hayashi}, {Hell}, {Hodges-kluck}, {Hornschemeier}, {Ichinohe}, {Ishi}, {Ishida}, {Ishikawa}, {Ishisaki}, {Kaastra}, {Kallman}, {Kara}, {Katsuda}, {Kanemaru}, {Kelley}, {Kilbourne}, {Kitamoto}, {Kobayashi}, {Kohmura}, {Kubota}, {Leutenegger}, {Loewenstein}, {Maeda}, {Markevitch}, {Matsumoto}, {Matsushita}, {Mccammon}, {Mcnamara}, {Mernier}, {Miller}, {Miller}, {Mitsuishi}, {Mizumoto}, {Mizuno}, {Mori}, {Mukai}, {Murakami}, {Mushotzky}, {Nakajima}, {Nakazawa}, {Ness}, {Nobukawa}, {Nobukawa}, {Noda}, {Odaka}, {Ogawa}, {Ogorzalek}, {Okajima}, {Ota}, {Paltani},
  {Petre}, {Plucinsky}, {Porter}, {Pottschmidt}, {Sato}, {Sato}, {Sawada}, {Seta}, {Shidatsu}, {Simionescu}, {Smith}, {Suzuki}, {Szymkowiak}, {Takahashi}, {Takeo}, {Tamagawa}, {Tamura}, {Tanaka}, {Tanimoto}, {Tashiro}, {Terada}, {Terashima}, {Tsuboi}, {Tsujimoto}, {Tsunemi}, {Tsuru}, {Uchida}, {Uchida}, {Uchida}, {Uchiyama}, {Ueda}, {Uno}, {Vink}, {Watanabe}, {Williams}, {Yamada}, {Yamada}, {Yamaguchi}, {Yamaoka}, {Yamasaki}, {Yamauchi}, {Yamauchi}, {Yaqoob}, {Yoneyama}, {Yoshida}, {Yukita}, {Zhuravleva}, {Seppi}, {Aihara}, \& {Omiya}}]{xrism.a2319.2025}
---. 2025{\natexlab{b}}, arXiv e-prints, arXiv:2508.05067, \dodoi{10.48550/arXiv.2508.05067}

\bibitem[{{Xrism Collaboration} {et~al.}(2025){Xrism Collaboration}, {Audard}, {Awaki}, {Ballhausen}, {Bamba}, {Behar}, {Boissay-Malaquin}, {Brenneman}, {Brown}, {Corrales}, {Costantini}, {Cumbee}, {Diaz Trigo}, {Done}, {Dotani}, {Ebisawa}, {Eckart}, {Eckert}, {Eguchi}, {Enoto}, {Ezoe}, {Foster}, {Fujimoto}, {Fujita}, {Fukazawa}, {Fukushima}, {Furuzawa}, {Gallo}, {Garc{\'\i}a}, {Gu}, {Guainazzi}, {Hagino}, {Hamaguchi}, {Hatsukade}, {Hayashi}, {Hayashi}, {Hell}, {Hodges-Kluck}, {Hornschemeier}, {Ichinohe}, {Ishi}, {Ishida}, {Ishikawa}, {Ishisaki}, {Kaastra}, {Kallman}, {Kara}, {Katsuda}, {Kanemaru}, {Kelley}, {Kilbourne}, {Kitamoto}, {Kobayashi}, {Kohmura}, {Kubota}, {Leutenegger}, {Loewenstein}, {Maeda}, {Markevitch}, {Matsumoto}, {Matsushita}, {McCammon}, {McNamara}, {Mernier}, {Miller}, {Miller}, {Mitsuishi}, {Mizumoto}, {Mizuno}, {Mori}, {Mukai}, {Murakami}, {Mushotzky}, {Nakajima}, {Nakazawa}, {Ness}, {Nobukawa}, {Nobukawa}, {Noda}, {Odaka}, {Ogawa}, {Ogorza{\l}ek}, {Okajima}, {Ota}, {Paltani}, {Petre},
  {Plucinsky}, {Porter}, {Pottschmidt}, {Sato}, {Sato}, {Sawada}, {Seta}, {Shidatsu}, {Simionescu}, {Smith}, {Suzuki}, {Szymkowiak}, {Takahashi}, {Takeo}, {Tamagawa}, {Tamura}, {Tanaka}, {Tanimoto}, {Tashiro}, {Terada}, {Terashima}, {Tsuboi}, {Tsujimoto}, {Tsunemi}, {Tsuru}, {T{\"u}mer}, {Uchida}, {Uchida}, {Uchida}, {Uchiyama}, {Ueda}, {Ueda}, {Uno}, {Vink}, {Watanabe}, {Williams}, {Yamada}, {Yamada}, {Yamaguchi}, {Yamaoka}, {Yamasaki}, {Yamauchi}, {Yamauchi}, {Yaqoob}, {Yoneyama}, {Yoshida}, {Yukita}, {Zhuravleva}, {Fabian}, {Nelson}, {Okabe}, {Pillepich}, {Potter}, {Regamey}, {Sakai}, {Shishido}, {Truong}, {Wik}, \& {Zuhone}}]{xrism.coma.2025}
{Xrism Collaboration}, {Audard}, M., {Awaki}, H., {et~al.} 2025, \apjl, 985, L20, \dodoi{10.3847/2041-8213/add2f6}

\bibitem[{{Yang} {et~al.}(2019){Yang}, {Gaspari}, \& {Marlow}}]{yang.etal.2019}
{Yang}, H. Y.~K., {Gaspari}, M., \& {Marlow}, C. 2019, \apj, 871, 6, \dodoi{10.3847/1538-4357/aaf4bd}

\bibitem[{{Zhuravleva} {et~al.}(2014){Zhuravleva}, {Churazov}, {Schekochihin}, {Allen}, {Ar{\'e}valo}, {Fabian}, {Forman}, {Sanders}, {Simionescu}, {Sunyaev}, {Vikhlinin}, \& {Werner}}]{zhuravleva.etal.2014}
{Zhuravleva}, I., {Churazov}, E., {Schekochihin}, A.~A., {et~al.} 2014, \nat, 515, 85, \dodoi{10.1038/nature13830}

\bibitem[{{Zinger} {et~al.}(2020){Zinger}, {Pillepich}, {Nelson}, {Weinberger}, {Pakmor}, {Springel}, {Hernquist}, {Marinacci}, \& {Vogelsberger}}]{zinger.etal.2020}
{Zinger}, E., {Pillepich}, A., {Nelson}, D., {et~al.} 2020, \mnras, 499, 768, \dodoi{10.1093/mnras/staa2607}

\bibitem[{{ZuHone} {et~al.}(2014){ZuHone}, {Biffi}, {Hallman}, {Randall}, {Foster}, \& {Schmid}}]{zuhone.etal.2014}
{ZuHone}, J.~A., {Biffi}, V., {Hallman}, E.~J., {et~al.} 2014, arXiv e-prints, arXiv:1407.1783, \dodoi{10.48550/arXiv.1407.1783}

\end{thebibliography}
\bibliographystyle{aasjournal}

\allauthors
\end{document}